\documentclass{pasj00}
\draft
\usepackage{color}

\begin{document}
\SetRunningHead{Author(s) in page-head}{Running Head}

\title{Distortion of Magnetic Fields in Barnard 68}

\author{Ryo \textsc{Kandori} %
}
\affil{Astrobiology Center of NINS, 2-21-1, Osawa, Mitaka, Tokyo 181-8588, Japan}
\email{r.kandori@nao.ac.jp}

\author{Motohide \textsc{Tamura}}
\affil{Astrobiology Center of NINS, 2-21-1, Osawa, Mitaka, Tokyo 181-8588, Japan / National Astronomical Observatory of Japan, 2-21-1 Osawa, Mitaka, Tokyo 181-8588, Japan / Department of Astronomy, The University of Tokyo, 7-3-1, Hongo, Bunkyo-ku, Tokyo, 113-0033, Japan}
\email{motohide.tamura@nao.ac.jp}

\author{Masao \textsc{Saito}}
\affil{National Astronomical Observatory of Japan, 2-21-1 Osawa, Mitaka, Tokyo 181-8588, Japan}
\email{masao.saito@nao.ac.jp}

\author{Kohji \textsc{Tomisaka}}
\affil{National Astronomical Observatory of Japan, 2-21-1 Osawa, Mitaka, Tokyo 181-8588, Japan}
\email{tomisaka@th.nao.ac.jp}

\author{Tomoaki \textsc{Matsumoto}}
\affil{Faculty of Sustainability Studies, Hosei University, Fujimi, Chiyoda-ku, Tokyo 102-8160}
\email{matsu@hosei.ac.jp}

\author{Nobuhiko \textsc{Kusakabe}}
\affil{Astrobiology Center of NINS, 2-21-1, Osawa, Mitaka, Tokyo 181-8588, Japan}
\email{nb.kusakabe@nao.ac.jp}

\author{Jungmi \textsc{Kwon}}
\affil{Department of Astronomy, The University of Tokyo, 7-3-1, Hongo, Bunkyo-ku, Tokyo, 113-0033, Japan}
\email{kwon.jungmi@astron.s.u-tokyo.ac.jp}

\author{Takahiro \textsc{Nagayama}}
\affil{Kagoshima University, 1-21-35 Korimoto, Kagoshima 890-0065, Japan}
\email{nagayama@sci.kagoshima-u.ac.jp}

\author{Tetsuya \textsc{Nagata}}
\affil{Kyoto University, Kitashirakawa-Oiwake-cho, Sakyo-ku, Kyoto 606-8502, Japan}
\email{nagata@kusastro.kyoto-u.ac.jp}

\author{Ryo \textsc{Tazaki}}
\affil{Astronomical Institute, Graduate School of Science Tohoku University, 6-3 Aramaki, Aoba-ku, Sendai 980-8578, Japan}
\email{rtazaki@astr.tohoku.ac.jp}

\and
\author{Ken'ichi {\sc Tatematsu}}
\affil{National Astronomical Observatory of Japan, 2-21-1 Osawa, Mitaka, Tokyo 181-8588, Japan}
\email{k.tatematsu@nao.ac.jp}

%


\maketitle

\begin{abstract}
The magnetic field structure, kinematical stability, and evolutionary status of the starless dense core Barnard 68 (B68) are revealed based on the near-infrared polarimetric observations of background stars, measuring the dichroically polarized light produced by aligned dust grains in the core. After subtracting unrelated ambient polarization components, the magnetic fields pervading B68 are mapped using 38 stars and axisymmetrically distorted hourglass-like magnetic fields are obtained, although the evidence for the hourglass field is not very strong. On the basis of simple 2D and 3D magnetic field modeling, the magnetic inclination angles on the plane-of-sky and in the line-of-sight direction are determined to be $47^{\circ} \pm 5^{\circ}$ and $20^{\circ} \pm 10^{\circ}$, respectively. The total magnetic field strength of B68 is obtained to be $26.1 \pm 8.7$ $\mu {\rm G}$. The critical mass of B68, evaluated using both magnetic and thermal/turbulent support, is  $M_{\rm cr} = 2.30 \pm 0.20$ ${\rm M}_{\odot}$, which is consistent with the observed core mass of $M_{\rm core}=2.1$ M$_{\odot}$, suggesting nearly critical state. We found a relatively linear relationship between polarization and extinction up to $A_V \sim 30$ mag toward the stars with deepest obscuration. Further theoretical and observational studies are required to explain the dust alignment in cold and dense regions in the core. 
\end{abstract}

\section{Introduction}
Barnard 68 (hereafter B68) is one of the most well defined low-mass starless dense cores. Due to its nearby location, isolated geometry, and rich stellar field lying behind the core, studies of the mass distribution of B68 has been made by near-infrared (NIR) extinction mapping toward background stars. NIR extinction studies of B68 by Alves et al. (2001b) show a density structure of the core that is remarkably well fitted by the Bonnor--Ebert model (Ebert 1955; Bonnor 1956). The core's physical properties were determined to be: Bonnor--Ebert solution parameter $\xi_{\rm max}=6.9\pm0.2$ (center-to-edge density contrast of $\rho_{\rm c}/\rho_{\rm s}=16.5$), radius $R = 100''=12,500$ AU, mass $M_{\rm core} = 2.1$ $M_{\odot}$, and boundary pressure $P_{\rm s} = 2.5 \times 10^{-12}$ Pa (i.e., $1.81 \times 10^5$ K ${\rm cm}^{-3}$) at an assumed distance $d$ of 125 pc (de Geus 1989). \par
%
Far-infrared (FIR) Herschel observations have revealed a colder dust temperature in the inner part of the core (Nielbock et al. 2012), and the column density structure of the core has been fitted with the parameters consistent with the NIR extinction studies (Roy et al. 2014). Sub-millimeter observations suggest a decreasing dust temperature toward the center of the core (Bianchi et al. 2003). \par
Measurements of the kinematic temperature $T_{\rm k}$ of the core range from 8 K to 16 K (Bourke et al. 1995; Hotzel et al. 2002a,b; Lai et al. 2002,2003; Lada et al. 2003). The lower value of 8 K was estimated based on the C$^{18}$O ($J=1$--$0$) excitation temperature and Monte Carlo modeling (Hotzel et al. 2002a), but is not based on NH$_3$ observations. The latter four papers on the kinetic temperature measurements cited above report $T_{\rm k} = 10$--$11$ K. Alves et al. (2001b) used a value of $T_{\rm k}=16$ K (Bourke et al 1995), but this value may be too high (private communication to T. Bourke, noted in Lada et al. 2003). \par
The measurements of $T_{\rm k}$ enable the distance to B68 to be estimated, under the assumption that Bonnor--Ebert equilibrium is sustained. Note that in Bonnor--Ebert solutions, $d^{-1}T$ is constant (Lai et al. 2003), and this constraint on the temperature can similarly constrain the distance to the core. On the basis of the temperature measurements, Hotzel et al. (2002b) and Lai et al. (2003) suggested that the distance to B68 is $85$--$100$ pc. However, caution must be used in this kind of analysis, because the existence of Bonnor--Ebert equilibrium has not been verified. It is known that a Bonnor--Ebert-like density structure can appear in \lq \lq out of equilibrium'' cores produced in turbulent simulations (Ballesteros-Paredes et al. 2003). Moreover, a dynamically collapsing core also mimics the Bonnor--Ebert density solution (Kandori et al. 2005). Since the Bonnor--Ebert model ignores support from magnetic fields, the results of distance analysis can be changed if magnetic support is included in the estimation of the equilibrium. For these reasons, determining the distance based on the Bonnor--Ebert model and temperature measurements may not be appropriate. \par
On the basis of sensitive high-resolution molecular line observations, Lada et al. (2003) suggested that the outer layer of B68 is pulsating. The narrow line width and lack of supersonic infall motion, as well as the pulsation, are consistent with the conclusions given by Alves et al. (2001b). In the Bonnor--Ebert fitting by Alves et al. (2001b), the obtained Bonnor--Ebert solution $\xi_{\rm max} = 6.9$ exceeds the value of the stability criterion $\xi_{\rm max} = 6.5$, but shows a nearly critical value, suggesting that B68 is on the verge of collapse. Pulsation of the core is confirmed by other observations (Redman et al. 2006; Maret et al. 2007) and theoretical studies (Keto et al. 2006; Broderick et al. 2007). \par
The final piece of B68's physical parameter is the magnetic field structure. While previous observations suggest that B68 is in a state not far from equilibrium, the stability of the core is not known in terms of magnetic support. The mass of B68 is 2.1 $M_{\odot}$ at an assumed distance of 125 pc (de Geus 1989), with $T_{\rm k}=16$ K (Bourke et al. 1995). If we use a realistic $T_{\rm k}$ value of $10$--$11$ K (Hotzel et al. 2002a,b; Lai et al. 2003; Lada et al. 2003), the Bonnor--Ebert mass is less than the actual mass of the core. This is inconsistent with the rather static features of the core observed in radio molecular lines (Lada et al. 2003). There are two possible solutions: either B68 sits at a closer distance ($85$--$100$ pc) or it exhibits additional support, such as support by magnetic fields. 
\par
In the present study, wide-field background star polarimetry at NIR wavelengths was conducted for B68. The magnetic field structure on the sky plane was revealed using several tens of stars within the core radius. The total magnetic field strength was estimated using the Davis--Chandrasekhar--Fermi method (Davis 1951; Chandrasekhar \& Fermi, 1953) and three dimensional (3D) magnetic field modeling of the core. Given the magnetic field information, the stability of B68 was determined and the evolutionary status of the core is discussed. The polarization vs. extinction ($P$--$A$) relationship in B68 was determined, correcting for the following: (1) ambient polarization, (2) depolarization effect caused by a distorted magnetic field shape, and (3) line-of-sight magnetic inclination angle. The obtained linear $P$--$A$ relationship ensures that the observed polarizations reflect the overall magnetic field structure in the core. The linear relationship between the polarization and extinction also raises a question regarding the alignment of the dust grains in the cold and dense environments in starless molecular cloud cores. 


\section{Observations and Data Reduction}
We observed B68 using the $JHK_s$-simultaneous imaging camera SIRIUS (Nagayama et al. 2003) and its polarimetry mode SIRPOL (Kandori et al. 2006) on the IRSF 1.4-m telescope at the South African Astronomical Observatory (SAAO). IRSF/SIRPOL provides deep- and wide- ($7.\hspace{-3pt}'7 \times 7.\hspace{-3pt}'7$ with a scale of 0$.\hspace{-3pt}''$45 ${\rm pixel}^{-1}$) field polarization images at $JHK_s$ simultaneously, which makes SIRPOL one of the most powerful instrument for NIR polarization surveys. SIRPOL is a single beam polarimeter. The uncertainty of polarization degree due to sky variation during exposures is typically $0.3\%$ under the photometric condition. The uncertainty of the determination of the origin of polarization angle of the polarimeter (i.e., correction angle) is less than 3$^{\circ}$ (Kandori et al. 2006 and updates for Kusune et al. 2015). 
\par
We conducted observations of B68 on the nights of 2007 June 18 and 19 using the linear polarimetry mode of SIRPOL. Ten-second exposures at four half-waveplate angles (in the sequence $0^{\circ}$, $45^{\circ}$, $22.5^{\circ}$, and $67.5^{\circ}$) were performed at ten dithered positions (1 set). The total integration time was 3900 seconds (39 sets) per wave plate angle. The seeing during the observations was typically $\sim$$1.\hspace{-3pt}''3$ (2.8 pixels) in the $H$ band. 
\par
The observed data were reduced using the Interactive Data Language (IDL) software, following the same manner as described in Kandori et al. (2007) (flat-field correction with twilight flat frames, median sky subtraction, and frame combine after registration). The point sources having a peak intensity greater than $10 \sigma $ above the local sky background were detected on the Stokes $I$ image, and aperture polarimetry was carried out for the sources on each waveplate angle image ($I_{0}$, $I_{45}$, $I_{22.5}$, and $I_{67.5}$). The sources with photometric errors greater than 0.1 mag were rejected. The number of detected sources in the field of view is 5817, 7383, and 4479 in the $J$, $H$, and $K_s$ bands, respectively. The limiting magnitudes are 18.5, 18.0, and 17.2 mag in the $J$, $H$, and $K_s$ bands, respectively. The aperture radius was the same as the full width at half maximum (FWHM) of stars (3.1, 2.8, and 2.8 pixels in the $J$, $H$, and $K_s$ bands), and the sky annulus was set to 10 pixels with a 5-pixel width. A relatively small aperture was used to suppress flux contaminations from neighboring sources. We did not use psf-fitting photometry, because the goodness of fitting for the stars on each waveplate angle image can be the source of systematic error. Note that we confirmed that the use of small aperture size does not affect our important conclusions based on the analyses by changing the size of aperture. 
\par
The Stokes parameters for each star can be obtained using $I = (I_{0} + I_{45} + I_{22.5} + I_{67.5})/2$, $Q = I_{0} - I_{45}$, and $U = I_{22.5} - I_{67.5}$. The polarization degree $P$ and polarization angle $\theta $ were derived by $P = \sqrt{Q^2 + U^2}/I$ and $\theta = 0.5 {\rm atan}(U/Q)$. Since the polarization degree, $P$, is positive quantity, the derived $P$ values tend to be overestimated, especially for low $S/N$ sources. We corrected the bias using $P_{\rm db} = \sqrt{P^2 - \delta P^2}$ (Wardle \& Kronberg 1974). 
\par
In the present study, we discuss the results in the $H$ band, where dust extinction is less severe than in the $J$ band, and polarization efficiency is greater than in the $K_s$ band. 
\section{Results and Discussion}
\subsection{Distortion of Magnetic Fields}
Figure 1 shows the observed polarization vector map of B68 in the $H$ band. The stars with $P_H / \delta P_H \ge 4$ are shown. B68 is located at the center of the image as a dark opaque region. Toward B68, the strongest polarization vector is $P_H \sim 10$ \%, and a bending field structure can be seen from the north to south-west direction. Outside B68 ($R > 100''$), weak but roughly uniform polarization vectors are distributed in the north-east to south-west direction. These polarization components can be regarded as ``off-core'' polarizations, located toward the same line of sight but unrelated to the B68 core. 
As shown in Figures 2 and 3, the off-core vectors have $P_{H,{\rm off}} = 1.1 \pm 0.36$ \% and $\theta_{H,{\rm off}} = 60^{\circ} \pm 16^{\circ}$ (median value and standard deviation). 
Following the same manner described in our previous paper (Kandori et al. 2017a, hereafter Paper I), we fitted the off-core vectors on the sky plane by $Q/I$ and $U/I$. The distributions of the $Q/I$ and $U/I$ values are modeled as $f(x,y)=A + Bx + Cy$, where $x$ and $y$ are the pixel coordinates, and $A$, $B$, and $C$ are the fitting parameters. 
\par
The estimated off-core vectors are shown in Figure 4. The (regression) off-core vectors were subtracted from the original vectors to isolate the polarization vectors associated with B68 (Figure 5). After subtraction, the polarization degree of the off-core vectors was successfully suppressed toward $0\%$ (Figure 2), and the polarization degree became roughly randomly distributed (Figure 3). The suppression of the off-core vectors can be confirmed in Figure 5. In Figure 5, a bending polarization vector structure can clearly be seen, particularly in the north-west half of the core. Though the bending structure is not very clear in the south-east half of the core, the direction of the vectors is slightly but systematically different from those in the north-west part. The number of polarization vectors within the core radius of $R \le 100''$ is 64. This number is not large, but is sufficient to trace the magnetic field lines. 

\subsection{Parabolic Modeling}
The most probable shape of the magnetic field lines, estimated using a parabolic function and its rotation, is shown in Figures 6 and 7 (solid white lines). The field of view is the same as the diameter of B68 ($200''$), and 43 polarization vectors with $P_H \ge 1$ \% are shown in the figure. The threshold of $P_H \ge 1$ \% was used, because the most of residual polarizations after the subtraction analysis is less than $1\%$ as shown in Figure 2. The threshold is set in order to avoid the systematic error caused by the subtraction analysis of off-core vectors. 
Figure 7 is the same as Figure 6 but the background image is a contour map of column density ($N_{\rm H_2}$) obtained with {\it Herschel} satellite (Roy et al. 2014). The contour was drawn in a step of $1.0 \times 10^{21}$ cm$^{-2}$. The minimum and maximum values are $1.6 \times 10^{21}$ cm$^{-2}$ and $1.0 \times 10^{22}$ cm$^{-2}$, respectively. The resolution of the map is the same as the SPIRE 500 $\mu$m data ($36.\hspace{-3pt}''3$). 
%
The five blue vectors located in the north-east part of the core are masked in the fitting, thus the number of vectors available for the fitting is 38. 
The coordinate origin of the parabolic function is fixed to the central position of the core measured on the column density map (R.A.=17$^{\rm h}$22$^{\rm m}$38$.\hspace{-3pt}^{\rm s}$6, Decl.=$-$23$^{\circ}$49$'$51$.\hspace{-3pt}''0$, J2000; Nielbock et al., 2012). \par
The best-fit parameters are $\theta_{\rm mag}=47^{\circ} \pm 5^{\circ}$ and $C = 2.20 (\pm 2.03) \times 10^{-5}$ ${\rm pixel}^{-2}$ ($= 1.09 \times 10^{-4}$ ${\rm arcsec}^{-2}$) for the parabolic function $y = g + gC{x^2}$, where $g$ specifies the magnetic field lines, $\theta_{\rm mag}$ denotes the position angle of the magnetic field direction (from north through east), and $C$ determines the degree of curvature in the parabolic function. 
The observational error of each star was taken into account in the calculations of $\chi^2 = (\sum_{i=1}^n (\theta_{\rm obs,{\it i}} - \theta_{\rm model}(x_i,y_i))^2 / \delta \theta_i^2$, where $n$ is the number of stars, $x$ and $y$ are the coordinates of the stars, $\theta_{\rm obs}$ and $\theta_{\rm model}$ denote the polarization angle from the observations and the model, and $\delta \theta_i$ is the observational error) in the fitting procedure. 
The existence of the hourglass field seems real, since the standard deviation of the residual angles, $\theta_{\rm res} = \theta_{\rm obs} - \theta_{\rm fit}$, is smaller for the parabolic function ($\delta \theta_{\rm res} = 15.63^{\circ} \pm 0.88^{\circ}$, Figure 8) than for the uniform field case ($19.75^{\circ} \pm 2.75^{\circ}$). \par
Note that the orientation of the five blue masked vectors shown in Figures 6 and 7 is roughly perpendicular to the best-fit magnetic field lines (while the other 38 vectors show no large dispersion in polarization angle with respect to the hourglass-fit lines). Though the number of blue vectors is small, these five vectors will thus affect the $\chi^2$ result if they are not removed. If we include the five blue vectors in the analysis, $\chi^2$ converges to the solution of a uniform field. We conclude that the finding of the hourglass-like magnetic field in B68 is not very strong. The five blue vectors locate in the periphery of the optical boundary of B68, and we expect that the mass (and magnetic field) distribution of the core at the region is not symmetric which can affect the obtained orientation of integrated polarization toward the line of sight. Thus, the reasons to mask the five blue vectors are 1) the number is small, 2) only they have large separations in polarization angle from the hourglass field lines, and 3) they are located at the periphery of the core.
\par
B68 is the second starless core to be fitted with an axisymmetrically distorted hourglass-like magnetic field, although the evidence is not strong compared with the case of FeSt 1-457 (Paper I). 
For comparison with FeSt 1-457, the smaller angular size of B68 causes the limited number of polarization data points, so that the shape of hourglass-like structure could not be well resolved. Moreover, the boundary shape of B68 is not circular. Thus, the mass distribution of the core may also deviate from symmetry toward the line of sight. These effects probably make the hourglass magnetic fields pervading B68 unclear. 
The obtained magnetic curvature value $C$ is similar to that exhibited for FeSt 1-457 ($C = 5.14 \times 10^{-5}$ ${\rm arcsec}^{-2}$, Paper I), suggesting the existence of similar mechanisms creating distortion in starless cores. 
\par
In Figure 5, there are several stars outside the core radius, to the Northeast, that follow the same distorted shape of magnetic fields shown within the radius area. This is not strange, because the distorted pattern of magnetic field lines is expected to smoothly connect to the outer magnetic fields surrounding the core (e.g., Myers et al. 2018). The polarization vectors toward the B68 core are contaminated with the foreground and background polarization, and we used the \lq \lq off-core'' region as the fitting part to estimate the polarization component that is unrelated to the core. The Northeast outer-core vectors were included in the fitting to estimate off-core polarization fields, and therefore we did not make use of these vectors in the analysis of hourglass-like field. 
\par
As described in Paper I, curved magnetic fields surrounding the core may be a byproduct of core formation. The magnetic field lines can be dragged by accumulating medium under the flux freezing condition, and this process naturally explains the formation of hourglass-like field geometry (solved analytically by theoretical studies: Mestel 1966; Ewertowski \& Basu 2013; Myers et al. 2018). 
\par
Under the assumption of frozen-in magnetic fields, the intrinsic dispersion of the magnetic field direction, $\delta \theta_{\rm int}$, can be attributed to the Alfv\'{e}n wave perturbed by turbulence. The intrinsic dispersion of the residual angle, $\delta \theta_{\rm int} = (\delta \theta_{\rm res}^2 - \delta \theta_{\rm err}^2)^{1/2}$, estimated using the parabolic fitting, is $14.72^{\circ} \pm 0.93^{\circ}$ (0.257 radian), where $\delta \theta_{\rm err}$ is the standard deviation of the observational error in the polarization measurements. The strength of the plane-of-sky magnetic field (${B}_{\rm pos}$) can be estimated from the relation ${B}_{\rm pos} = {C}_{\rm corr} (4 \pi \rho)^{1/2} \sigma_{\rm turb} / \delta \theta_{\rm int}$, where $\rho$ and $\sigma_{\rm turb}$ are the mean density of the core and turbulent velocity dispersion (Davis 1951; Chandrasekhar \& Fermi, 1953), and ${C}_{\rm corr}=0.5$ is a correction factor suggested by theoretical studies (Ostriker et al., 2001, see also, Padoan et al. 2001; Heitsch et al. 2001; Heitsch 2005; Matsumoto et al. 2006). Using the data of mean density $(\rho = 1.53 (\pm 0.16) \times 10^{-19}$ g cm$^{-3}$) calculated from Alves et al. (2001b) and turbulent velocity dispersion $(\sigma_{\rm turb} = 0.0909$ km s${}^{-1}$ by using the line width of $\Delta V_{\rm N_2 H^+ ({\it J}=1-0)} = 0.25$ km s${}^{-1})$ taken from Lada et al. (2003) and $\delta \theta_{\rm int}$ derived in the present study, a relatively weak magnetic field is obtained as a lower limit of the total field strength ($|B|$): $B_{\rm pos} = 24.5$ $\mu {\rm G}$. 
For error estimate, we assumed that the value of turbulent velocity dispersion is accurate within 30\%. Then, the uncertainty of $B_{\rm pos}$ is estimated to be $8.2$ $\mu {\rm G}$.
\par
Note that we did not employ the method based on structure function (Hildebrand et al. 2009; Houde et al. 2009) to estimate $\delta \theta_{\rm int}$. It is not known that the combination of hourglass field and uniform field can be separated well on the structure function. This should be investigated with detailed model calculations before applying to the observational data. Though such investigation is beyond the scope of this paper, we plan to check whether the structure function method is available to the hourglass magnetic field data. 
\subsection{3D Magnetic Field}
For 3D magnetic field modeling, we follow the procedure described in a previous paper (Kandori et al. 2017b, hereafter Paper II, see also, Kandori et al. 2019, hereafter Paper VI). The 3D version of the simple parabolic function employed in Paper I, $z(r, \varphi, g) = g + gC{r}^{2}$ in cylindrical coordinates $(r, z, \varphi)$, is used to model the core magnetic fields, where $g$ specifies the magnetic field line, $C$ is the curvature of the lines, and $\varphi$ is the azimuth angle (measured in the plane perpendicular to the $r$). In the function, the shape of the magnetic field lines is axially symmetric around the $r$ axis. 
\par
The 3D model was virtually observed after rotating in the line of sight ($\gamma_{\rm mag}$) and the plane of sky ($\theta_{\rm mag}$) directions. For the analysis, we followed the same manner described in Section 3.1 of Paper VI (see also, Sections 2 and 3.1 of Paper II). The resulting polarization vector maps of the 3D parabolic model are shown in Figure 9. The white line shows the polarization vector, and the background color and color bar show the polarization degree of the model core. The density structure of the core was assumed to be the same as the Bonnor--Ebert sphere with the solution parameter of 6.9. The 3D magnetic curvature was set to $C = 2.0 \times 10^{-4}$ arcsec$^{-2}$ for all the panels. The applied viewing angle $\gamma_{\rm view} = 90^{\circ} - \gamma_{\rm mag}$, i.e., the angle between the line of sight and the magnetic axis, is labeled in the upper left corner of each panel. 
\par
The model polarization vector maps change depending on the viewing angle ($\gamma_{\rm view}$). As described in Paper II and VI, there are four characteristics: 1) decrease of maximum polarization degree from $\gamma_{\rm view} = 90^{\circ}$ to $\gamma_{\rm view} = 0^{\circ}$, 2) hourglass-shaped polarization angle pattern in large $\gamma_{\rm view}$ converges to radial pattern toward small $\gamma_{\rm view}$, 3) depolarization occurs in the polarization vector map, especially along the equatorial plane of the core, and 4) elongated structure of polarization degree distribution toward small $\gamma_{\rm view}$. The model map can be compared with observations for various inclination angle $\gamma_{\rm mag}$. Since polarization distributions of the model cores are different from each other, depending on the inclination angle $\gamma_{\rm mag}$ (Figure 9), $\chi^2$ fitting of those with the observational data can be used to restrict the line of sight magnetic inclination angle of the core. 
\par
Figure 10 shows the distribution of $\chi^2_\theta = (\sum_{i=1}^n (\theta_{\rm obs,{\it i}} - \theta_{\rm model}(x_i,y_i))^2 / \delta \theta_i^2$, where $n$ is the number of stars, $x$ and $y$ are the coordinates of the stars, $\theta_{\rm obs}$ and $\theta_{\rm model}$ denote the polarization angle from the observations and the model, and $\delta \theta_i$ is the observational error) calculated using the model and observed polarization angles. The best magnetic curvature parameter, $C$, was determined at each inclination angle $\gamma_{\rm mag}$ to obtain $\chi^2_\theta$. 
From Figure 10, the best $\gamma_{\rm mag}$ for B68 is $20^{\circ}$. The $\chi^2_\theta$ value becomes large toward pole-on geometry of magnetic fields. 
\par
To check consistency of the $\chi^2$ analysis based on polarization angle, we made $\chi^2$ calculations using both polarization angles and degrees. 
We determined the best model parameters for each $\gamma_{\rm mag}$ by minimizing the difference in polarization angles, and we calculated $\chi^2_P = (\sum_{i=1}^n (P_{\rm obs,{\it i}} - P_{\rm model}(x_i,y_i))^2 / \delta P_i^2$, where $P_{\rm obs}$ and $P_{\rm model}$ show the polarization degree from observations and the model, and $\delta P_i$ is the observational error) in polarization degrees (Figure 11). In the procedure, the relationship between the model core column density and polarization degree was scaled to be consistent with observations. 
In Figure 11, the minimization point is $\gamma_{\rm mag} = 20^{\circ}$, resulting in the same angle as the $\chi^2$ analysis based on polarization angle (Figure 10). The distributions of $\chi^2_{\theta}$ and $\chi^2_{P}$ show large scatter in the region of $\gamma_{\rm mag} \ge 80^{\circ}$. This can be due to the two effects. One is the polarization distribution of the model core which changes dramatically in the region near the pole-on view geometry. Another is the relatively small number of data points ($N=38$) of the B68's background stars. 
\par
The 1-sigma error estimated at the minimum $\chi^2_{\theta}$ point is $4.8^{\circ}$ and at the minimum $\chi^2_{P}$ point is $6.9^{\circ}$. Judging from the relatively flat $\chi^2_{\theta}$ and $\chi^2_{P}$ distribution especially from $\gamma_{\rm mag} = 10^{\circ}$ to $30^{\circ}$, we conclude that the magnetic inclination angle of B68 can be $\gamma_{\rm mag} = 20^{\circ} \pm 10^{\circ}$. The magnetic curvature obtained at $\gamma_{\rm mag} = 20^{\circ}$ is $C = 1.19 \times 10^{-3}$ ${\rm arcsec}^{-2}$. 
\par
Note that the $\chi^2$ values in Figures 10 and 11 are relatively large. This seems to be due to the existence of polarization angle scatter mainly caused by the Alfv\'{e}n wave, which cannot be included in the term of observational error in the calculation of $\chi^2$. 
\par
%
Figure 12 shows the best-fit 3D parabolic model with the observed polarization vectors. The direction of the model polarization vectors generally agree with observations, although several vectors seem to deviate from the model field. 
The standard deviation of the angular difference in the plane-of-sky polarization angles between the 3D model and observations is $9.45^{\circ}$, which is less than the fitting result in 2D as well as the uniform field case of $19.75^{\circ}$. 
\par
Figure 13 is the same observational data as Figure 12, but with the background image processed using the line integral convolution technique (LIC: Cabral \& Leedom 1993). We used publicly available interactive data language (IDL) code developed by Diego Falceta-Gon\c{c}alves. The direction of the LIC \lq \lq texture'' is parallel to the direction of magnetic fields, and the background image is based on the polarization degree of model core.

\subsection{Stability and Evolutionary Status of the Core}
On the basis of the obtained magnetic inclination angle $\gamma_{\rm mag}$ of $20^{\circ} \pm 10^{\circ}$, the total magnetic field strength of B68 is determined to be $B_{\rm pos}/\cos (\gamma_{\rm mag})=24.5/\cos(20^{\circ})=26.1 \pm 8.7$ $\mu{\rm G}$. 
The magnetic support of the core against gravity can be investigated using the parameter ${\lambda} = ({M}/{\Phi})_{\rm obs} / ({M}/{\Phi})_{\rm critical}$, which represents the ratio of the observed mass-to-magnetic flux ratio to a critical value $(2\pi {\rm G}^{1/2})^{-1}$ suggested by theory (Mestel \& Spitzer 1956; Nakano \& Nakamura 1978). 
We found $\lambda = 2.37 \pm 0.21$. The magnetic critical mass of the core, $M_{\rm mag} = 0.89 \pm 0.12$ ${\rm M}_{\odot}$, is clearly lower than the observed core mass, $M_{\rm core}=2.1$ ${\rm M}_{\odot}$, suggesting magnetically supercritical state. 
%
Though B68 is found to be magnetically supercritical, this does not necessarily imply an unstable state of the core. Including the contributions from magnetic and thermal/turbulent support, the critical mass of the core can be $M_{\rm cr} \simeq M_{\rm mag}+M_{\rm BE}$ (Mouschovias \& Spitzer 1976; Tomisaka, Ikeuchi, \& Nakamura 1988; McKee 1989), where $M_{\rm BE}$ is the Bonnor--Ebert mass. 
For B68, $M_{\rm cr}$ is $0.89+1.41=2.30 \pm 0.20$ ${\rm M}_{\odot}$, where $1.41 \pm 0.16$ ${\rm M}_{\odot}$ is the Bonnor--Ebert mass calculated using the kinematic temperature of the core of 10.5 K (Hotzel et al. 2002a,b; Lai et al. 2003; Lada et al. 2003), turbulent velocity dispersion of $0.0909$ ${\rm km s}^{-1}$ (Lada et al. 2003), and external pressure $P_{\rm ext}$ of $1.81(\pm 0.13) \times 10^5$ K cm$^{-3}$. 
The obtained $M_{\rm cr}$ is coincident to the observed core mass $M_{\rm core}$, and we thus conclude that the stability of B68 is in a condition close to the critical state. 
%
\par
It is now known that B68 is a critical core, located in a state just before or after the onset of dynamical collapse to form stars. The critical characteristics of the core are consistent with (1) the starless characteristic of the core, (2) the lack of detection of supersonic infalling gas motion (Lada et al. 2003), and (3) the detection of possible oscillating gas motion in the outer layer of the core (Lada et al. 2003). Further magnetic diffusion and/or turbulent dissipation can eventually trigger the collapse of the core. Note that the influence from the rotation of the core is small. The ratio of rotational energy to gravitational energy in the core is only 0.04 (Lada et al. 2003). 
\par
It is noteworthy that the B68 core can reach near equilibrium state at the employed distance of 125 pc and temperature of $10-11$ K, with the aids from magnetic fields. As stated in Introduction, if we consider thermal/turbulent support alone, the core readily starts collapse, which does match the observed static features of the core. The extra support from magnetic fields can naturally explain the physical status of B68, rather than assuming closer distance for the core. 
\par
%
%
The relative importance of magnetic fields in the support of the core is investigated based on the ratio of the thermal and turbulent energy to the magnetic energy, ${\beta} \equiv 3{C}_{\rm s}^{2}/{V}_{\rm A}^{2}$ and ${\beta}_{\rm turb} \equiv {\sigma}_{\rm turb,3D}^{2}/{V}_{\rm A}^{2} = 3{\sigma}_{\rm turb,1D}^{2}/{V}_{\rm A}^{2}$, where ${C}_{\rm s}$, ${\sigma}_{\rm turb}$, and ${V}_{\rm A}$ denote the isothermal sound speed at 10.5 K, turbulent velocity dispersion, and Alfv\'{e}n velocity. 
These ratios were found to be ${\beta} \sim 3.15$ and ${\beta}_{\rm turb} \sim 0.70$. B68 is dominated by thermal support with a smaller contribution from static magnetic fields. Turbulence seems to be dissipated but has a comparable contribution to that of magnetic fields. 
\par
Figure 14 shows the relationship between the core's elongation axis ($\theta_{\rm elon} \sim 140^{\circ}$), rotation ($\theta_{\rm rot}=23.5^{\circ}$, Lada et al. 2003: this value is not the direction of the velocity gradient but is the orientation of the rotation axis --- private communication with Lada), and magnetic field ($\theta_{\rm mag}=47^{\circ}$) superimposed on the map of dust extinction (Alves et al. 2001a). The rotation axis is taken from the velocity gradient fit of N$_2$H$^+$ $(J=1$--$0)$ observations (Lada et al. 2003). Though Lada et al. (2003) has reported a similar fit for C$^{18}$O $(J=1$--$0)$ line, significant molecular depletion is reported for this molecule (Bergin et al. 2002; Di Francesco et al. 2002; Hotzel et al. 2002a). We thus use the value obtained in N$_2$H$^+$. 
\par
As for FeSt 1-457, the orientation of the elongated structure in the opaque part of B68 is clearly perpendicular to the magnetic axis. This geometrical relationship is consistent with the picture of mass accretion along magnetic field lines suggested by theories (e.g., Galli \& Shu 1993a,b) or the magneto-hydrostatic configuration (e.g., Tomisaka, Ikeuchi, \& Nakamura 1988). A magnetohydrostatic configuration can produce a flattened inner density distribution whose major axis is perpendicular to the direction of the magnetic fields (Tomisaka, Ikeuchi, \& Nakamura 1988). This is consistent with the observed magnetic geometry and extinction distribution of the core. 
\par

The ratio $M_{\rm mag} / M_{\rm BE}$ can affect the shape of density distribution of magnetohydrostatic core. A large ratio corresponds to the magnetically dominated case, and a small ratio corresponds to the thermally supported case (the case, $M_{\rm mag} / M_{\rm BE} = 0$, is the Bonnor--Ebert sphere). The magnetohydrostatic cores can become elongated with increasing $M_{\rm mag} / M_{\rm BE}$. For FeSt 1-457, $M_{\rm mag} / M_{\rm BE} \approx 1.8$, whereas for B68  $M_{\rm mag} / M_{\rm BE} \approx 0.6$. The importance of magnetic support is different for these cores. The $A_V$ distribution (Figure 8 of Paper I) of FeSt 1-457 shows clear elongation around the center, whereas column density distribution (Figures 7 and 14) of B68 seems more circular around the center. For the two cores, a study based on detailed comparison between observed density and magnetic field structure with theory is planned. 
%
\par
There is a slight misalignment between the rotational axis and the magnetic axis. The initial angular momentum of the core was probably not very well aligned with the magnetic axis. Note that the rotational axis measured in C$^{18}$O $(J=1$--$0)$ line is in the north--south direction (Lada et al. 2003), although the direction of the axis in Figure 14 (N$_{2}$H$^{+}$) seems to be close to the magnetic axis of the core. \par
In the neighboring region of B68, there is a chain of small dark clouds, B69, B70, and B71. Including B68, they may have been created from the same filamentary cloud, and the magnetic axis of B68 is roughly perpendicular to the orientation of the chain, suggesting magnetically controlled structure formation. 
%

\subsection{Polarization$-$Extinction Relationship}
The relationship between dust polarization $P$ and extinction $A$ in dark clouds is important in terms of (1) interpreting the orientation of interstellar polarizations which is thought to be closely related to the direction of magnetic fields, and (2) investigating/constraining the mechanism of dust grain alignment with magnetic fields. The former point is important to ensure that the observed polarization can trace magnetically aligned dust in dense environments. If there are no linear relation (e.g., existence of kink) in $P$--$A$ relationship, observational interpretation of magnetic fields pervading dense cores becomes inaccurate. Determining accurate $P$--$A$ relationship for various cloud samples remains a problem for observational studies of dark clouds and cores. The latter point is important for comparing observations with theoretical models of dust grain alignment. These two points have been discussed in our previous paper using the data toward the starless dense core FeSt 1-457 (Kandori et al. 2018, hereafter Paper III, see also Paper VI). 
\par
The observed polarizations toward B68 are a superposition of the polarizations from the core and the ambient medium. The polarizations from the ambient off-core medium should be subtracted from the observed polarizations in order to isolate the polarization associated with the B68 core (Section 3.1). As described in Section 3.3, distorted magnetic fields surrounding the core can produce a depolarization effect. Furthermore, the line-of-sight magnetic inclination angle weakens the observed plane-of-sky polarizations. These effects should be corrected to obtain accurate $P$--$A$ relationship in B68.
\par
Figure 15(a) shows the observed $P_H$ vs. $H-K_{s}$ relationship with no correction. In the figure, the polarization generally increases with increasing extinction up to $H-K_{s} \sim 1.7$ mag, although there is a scatter in the distribution of the data points. The slope of the relationship is $3.11 \pm 0.11$ $\%$ ${\rm mag}^{-1}$. In Figure 15, the value of $A_V$ was added in each panel, which was calculated using the $A_V = 21.7 \times E_{H-K_s}$ relationship (Nishiyama et al. 2008). $E_{H-K_s}$ was obtained by subtracting average color of off-core stars, $<H-K_s>_{\rm off-core} = 0.34$ mag, from $H-K_s$ color of stars. 
\par
To subtract ambient off-core polarizations, the stars located outside the core radius, $R>100''$, are used to estimate the off-core polarization vectors. A relatively linear $P_H$ vs. $H-K_{s}$ relationship is obtained after subtracting the ambient polarization components, as shown in Figure 15(b). The slope of the relationship is $3.60 \pm 0.09$ $\%$ ${\rm mag}^{-1}$, which is similar to the value of the average interstellar polarization slope (Jones 1989). 
\par
%
%
B68 is associated with inclined distorted magnetic fields, as described and concluded in Sections 3.1--3.3, which provides depolarization effects inside the core. With known 3D magnetic field structure, the effect from core's depolarization and inclination can be corrected. In Figure 16, the distribution of correction factor is shown, and depolarization regions around the equatorial plane can clearly be seen. The figure was made based on a comparison of the polarization degree distribution of the inclined model core ($\gamma_{\rm mag} = 20^{\circ}$) with the model core with no inclination ($\gamma_{\rm mag} = 0^{\circ}$). Using the factor map, the effect from depolarization and inclination can be corrected simultaneously. 
\par
%
%
Figure 15(c) shows the final correction, the correction of depolarization effect and magnetic inclination angle. The Figure 15(b) relationship was divided by the correction factor map (Figure 16). 
The slope of the relationship is $4.57 \pm 0.11$ $\%$ ${\rm mag}^{-1}$, which is comparable to the value obtained for FeSt 1-457 ($6.60 \pm 0.41$ $\%$ ${\rm mag}^{-1}$, Paper VI). The polarization efficiencies for these cores, after geometric and depolarization corrections, are less than the statistically estimated upper limit for interstellar medium ($P_H / E_{H-K_s} \sim 14$, Jones 1989). 
Thus, there may be other factors to increase the polarization efficiency. For example, different dust properties and/or different radiation environment (radiative torque grain alignment theory: Dolginov \& Mitrofanov 1976; Draine \& Weingartner 1996,1997; Lazarian \& Hoang 2007) may play a role to determine the upper limit value. 
%
%
%
The correlation coefficient of the data points is 0.81, which is clearly larger than those of the original relationship of 0.65 and the Figure 15(b) relationship of 0.73, showing that our corrections have improved the tightness of the $P$--$A$ relationship. 
\par
%
%
%
Figure 17 shows the relationship between the polarization efficiency $P_H / A_V$ and $A_V$. The dashed line shows the linear least-squares fitting to the data points with $A_V > 7$ mag, resulted in $-0.0031 A_V + 0.3081$. The relatively linear relationship in Figure 15(c) is reflected in the shallow slope of the linear fitting result. The dotted line shows the fitting of the whole data using the power-law $P_H / A_V \propto A_V^{-\alpha}$, resulted in the $\alpha$ index of $0.34 \pm 0.12$. The shallow $\alpha$ index indicates that the polarization arisen in deep inside the core can be traced with our observations. The dotted-dashed line shows an observational upper limit by Jones (1989). The relation was calculated based on the equation $P_{K,{\rm max}} = \tanh{\tau_{\rm p}}$, where $\tau_{\rm p} = (1-\eta)\tau_{K}/(1+\eta)$, and the parameter $\eta$ is set to 0.875 (Jones 1989). $\tau_{K}$ denotes the optical depth in the $K$ band.
\par
%
For B68, we finally obtained a relatively linear relationship between the polarization and dust extinction up to $A_V \sim 30$ mag. The result ensures that the NIR polarimetric observations trace overall polarization (magnetic field) structure of B68. The alignment of dust grains inside cold and dense starless core remains a problem of astrophysics. Further observational and theoretical studies are desirable. 

\section{Summary and Conclusion}
The present study revealed the detailed magnetic field structure, kinematical stability, and evolutionary status of the starless dense core B68 based on NIR polarimetric observations of background stars measuring dichroically polarized light produced by aligned dust grains in the core. After subtracting unrelated ambient polarization components, the magnetic fields pervading B68 were mapped using 38 stars and axisymmetrically distorted hourglass-like magnetic fields were obtained, although the evidence for the hourglass field is not very strong. On the basis of simple 2D and 3D magnetic field modeling, the magnetic inclination angles on the plane-of-sky and in the line-of-sight direction were determined to be $47^{\circ} \pm 5^{\circ}$ and $20^{\circ} \pm 10^{\circ}$, respectively. Given the obtained magnetic inclination angle and using the Davis--Chandrasekhar--Fermi technique, the total magnetic field strength of B68 was obtained to be $B_{\rm pos}/\cos(\gamma_{\rm mag}) = 24.5/\cos(20^{\circ}) = 26.1 \pm 8.7$ $\mu$G. The magnetic critical mass of the core, $M_{\rm mag}=0.89 \pm 0.12$ M$_{\odot}$, is less than the observed core mass, $M_{\rm core}=2.1$ M$_{\odot}$, suggesting a magnetically supercritical state with the ratio of observed mass-to-magnetic flux to a critical value, $\lambda = 2.37 \pm 0.21$. The critical mass of B68, evaluated using both magnetic and thermal+turbulent support, $M_{\rm cr} \simeq M_{\rm mag}+M_{\rm BE}$, is $0.89+1.41=2.30 \pm 0.20$ ${\rm M}_{\odot}$, where $1.41 \pm 0.16$ ${\rm M}_{\odot}$ is the Bonnor--Ebert mass. We conclude that B68 is in a condition close to the critical state, with $M_{\rm cr} \sim M_{\rm core}$. The direction of the plane-of-sky magnetic fields of B68 ($47^{\circ}$) is obviously perpendicular to the core's elongation axis ($\sim 140^{\circ}$). This geometrical relationship is consistent with the magneto-hydrostatic configuration or the picture of mass accretion along magnetic field lines suggested by theories of isolated star formation. We found a linear relationship in the polarization vs. extinction diagram up to $A_V \sim 30$ mag toward the stars with deepest obscuration. The linear relationship indicates that the observed polarizations reflect the overall magnetic field structure of the core. Further theoretical and observational studies are necessary to explain the dust alignment in cold and dense regions in the core. 

\bigskip

We are grateful to the staff at SAAO for their kind help during the observations. We wish to thank Tetsuo Nishino, Chie Nagashima, and Noboru Ebizuka for their support in the development of SIRPOL and its calibration and its stable operation with the IRSF telescope. The IRSF/SIRPOL project was initiated and supported by Nagoya University, National Astronomical Observatory of Japan, and the University of Tokyo in collaboration with the South African Astronomical Observatory under the financial support of Grants-in-Aid for Scientific Research on Priority Area (A) No. 10147207 and No. 10147214, and Grants-in-Aid No. 13573001 and No. 16340061 of the Ministry of Education, Culture, Sports, Science, and Technology of Japan. RK, MT, NK, KT (Kohji Tomisaka), and MS also acknowledge support by additional Grants-in-Aid Nos. 16077101, 16077204, 16340061, 21740147, 26800111, 16K13791, 15K05032, 16K05303, and 19K03922.

\clearpage 

\begin{figure}[t]  
\begin{center}
 \includegraphics[width=6.5 in]{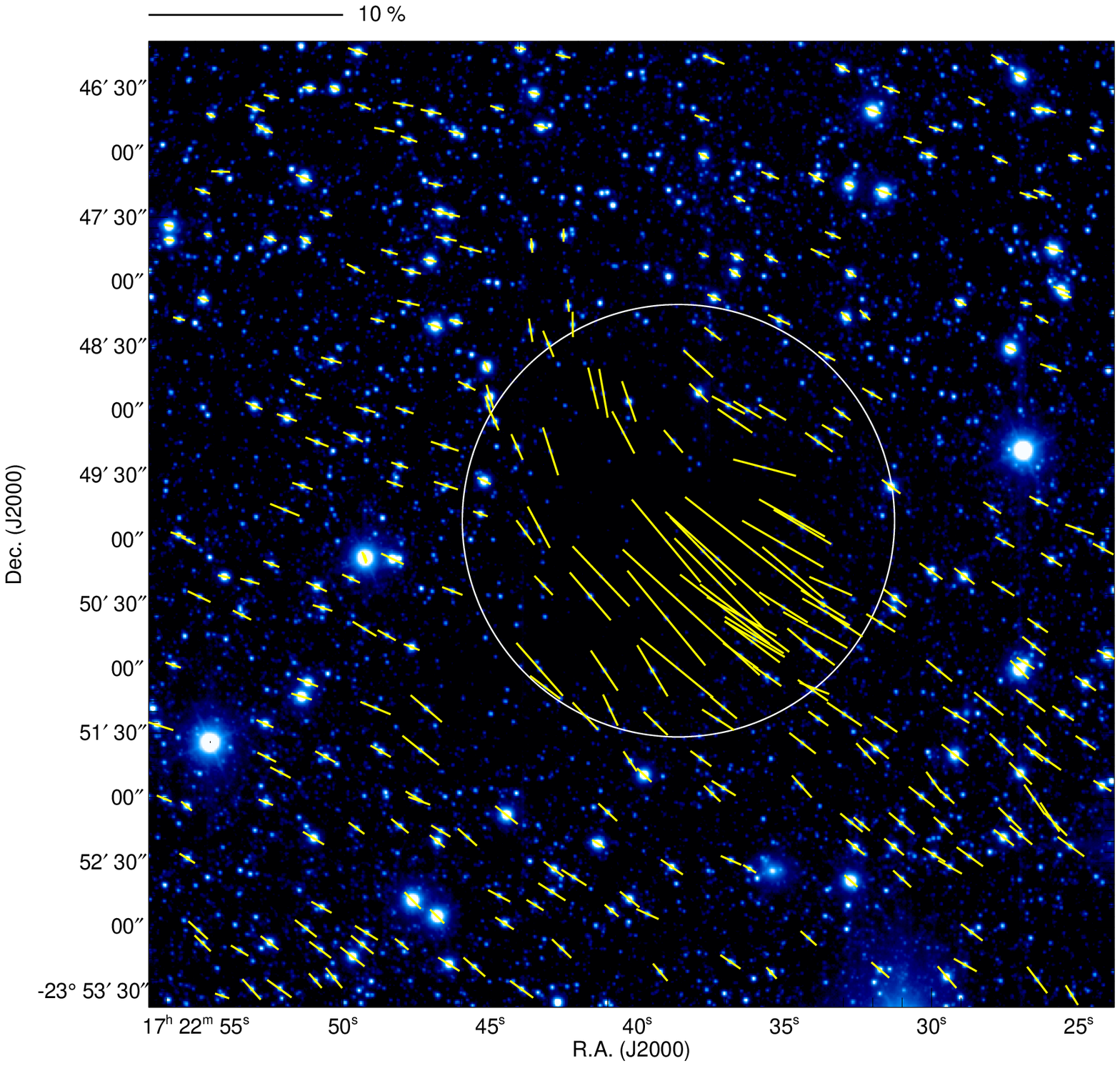}
\end{center}
 \caption{Polarization vectors of point sources superimposed on the intensity image in the $H$ band. The stars with $P_H / \delta P_H \ge 4$ are shown. The core radius (100$''$) is indicated by the white circle. The scale of the 10\% polarization degree is shown above the image.}
   \label{fig1}
\end{figure}

\clearpage 

\begin{figure}[t]  
\begin{center}
 \includegraphics[width=6.5 in]{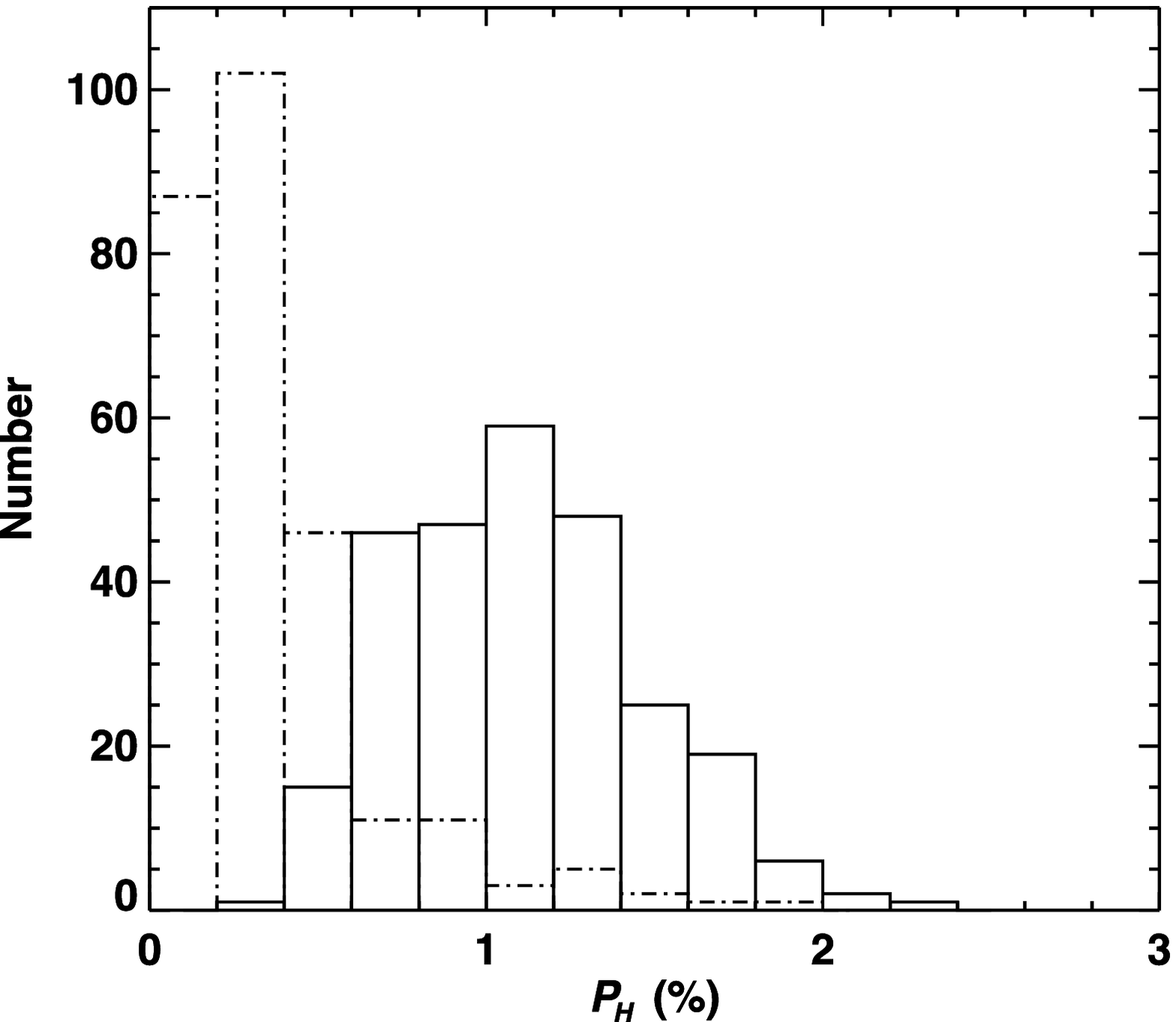}
\end{center}
 \caption{Histogram of $P_H$ for the stars in the off-core region before (solid line) and after (dotted-dashed line) subtraction of the off-core component.}
   \label{fig1}
\end{figure}

\clearpage 

\begin{figure}[t]  
\begin{center}
 \includegraphics[width=6.5 in]{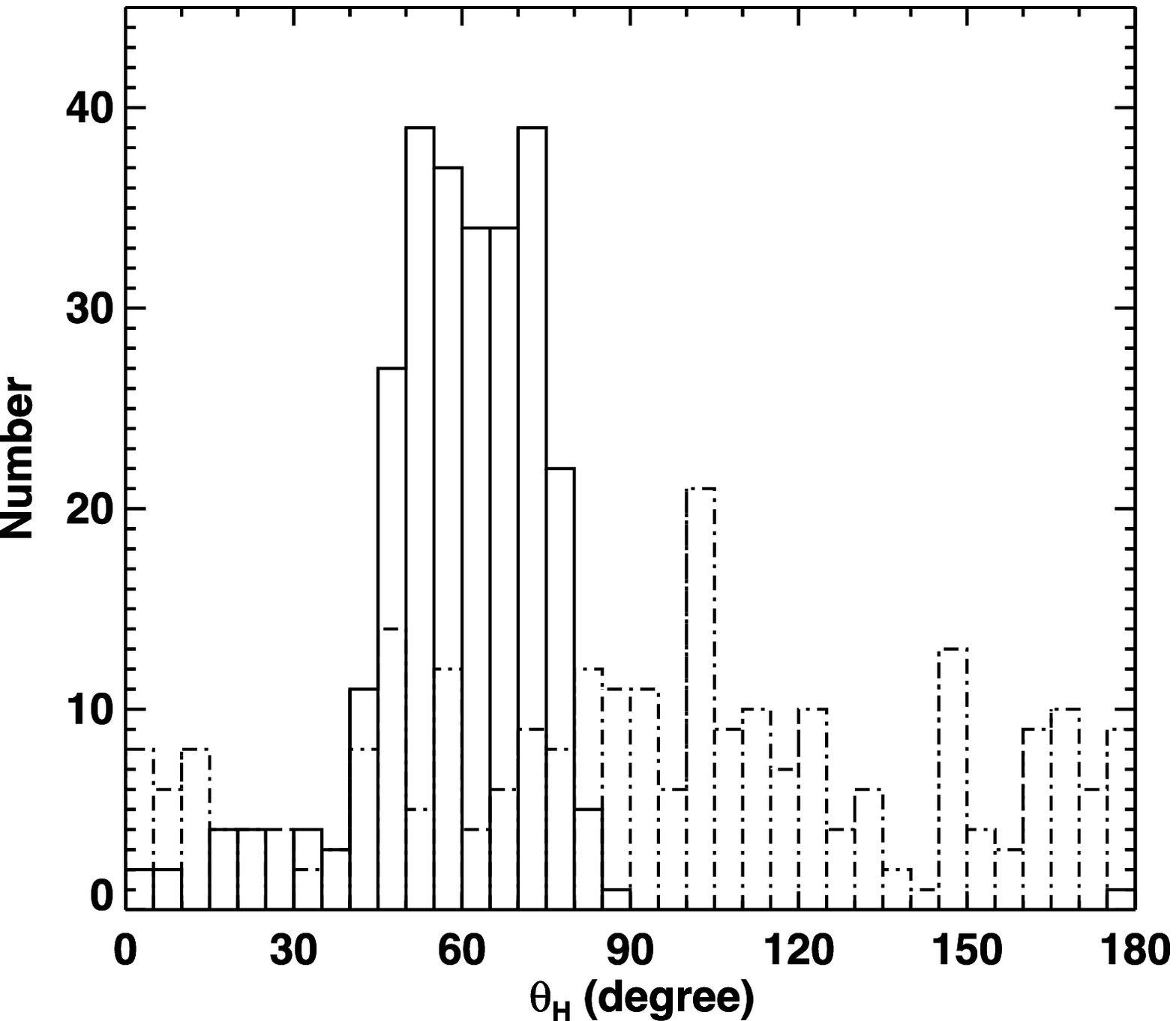}
\end{center}
 \caption{Histogram of $\theta_H$ for the stars in the off-core region before (solid line) and after (dotted-dashed line) subtraction of the off-core component.}
   \label{fig1}
\end{figure}

\clearpage 

\begin{figure}[t]  
\begin{center}
 \includegraphics[width=6.5 in]{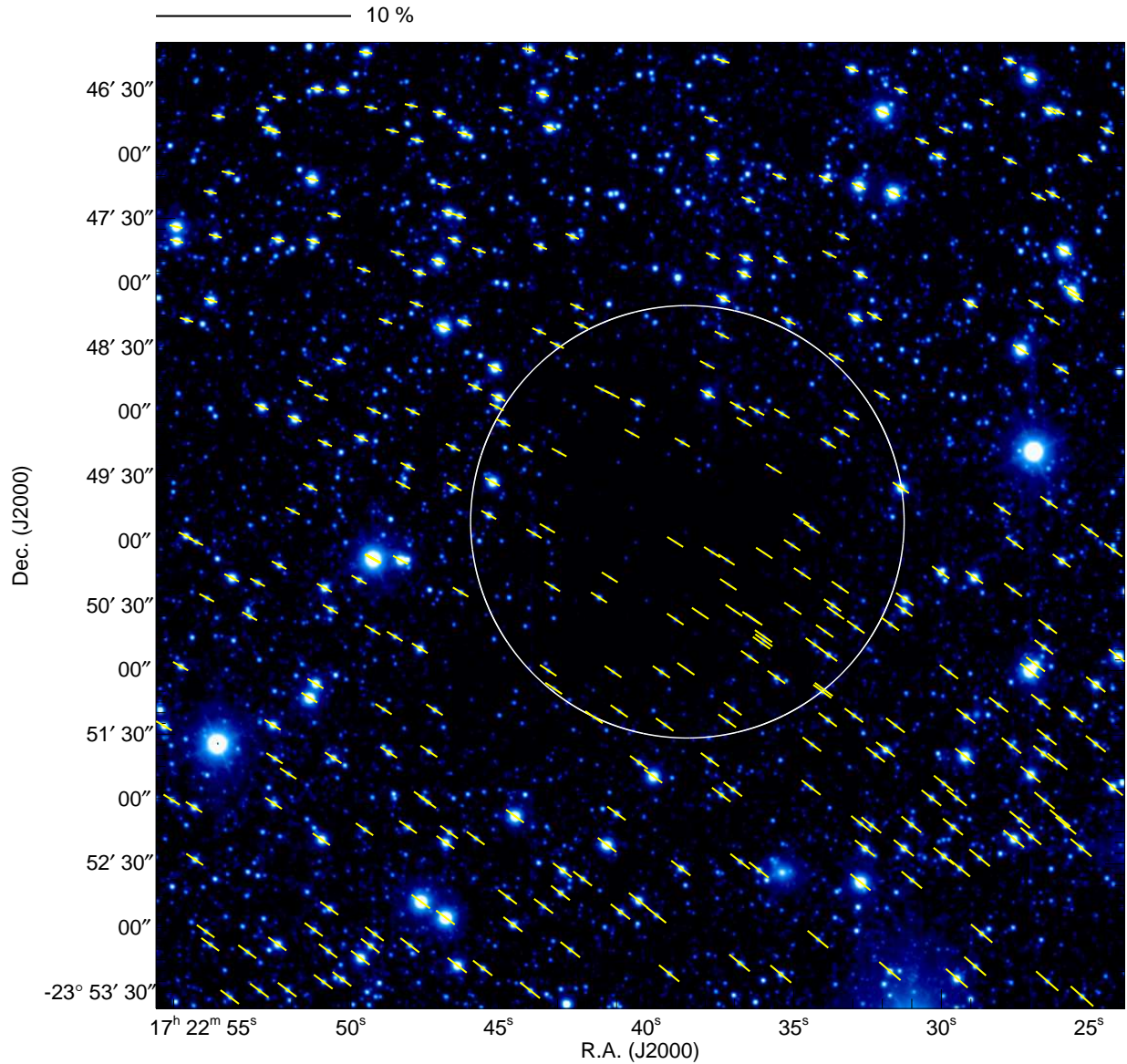}
\end{center}
 \caption{Estimated off-core polarization vectors superimposed on an intensity image in the $H$ band. Note that these vectors are not obtained directly from observations. Off-core vectors estimated by fitting are plotted at the position of each star. The scale of the 10\% polarization degree is shown above the image.}
   \label{fig1}
\end{figure}

\clearpage 

\begin{figure}[t]  
\begin{center}
 \includegraphics[width=6.5 in]{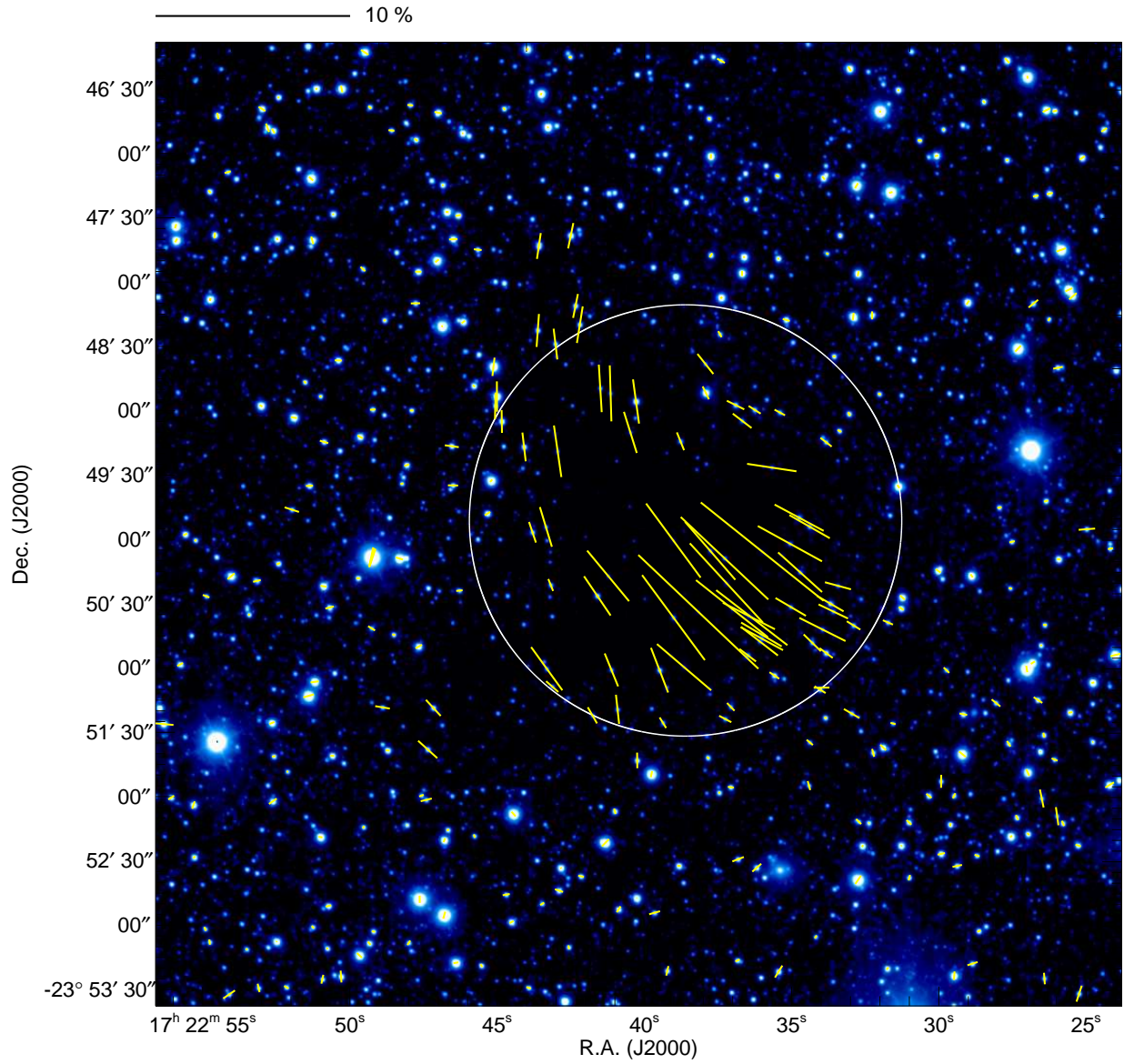}
\end{center}
 \caption{Polarization vectors after subtraction of the off-core component. The scale of the 10\% polarization degree is shown above the image.}
   \label{fig1}
\end{figure}

\clearpage 

\begin{figure}[t]  
\begin{center}
 \includegraphics[width=6.5 in]{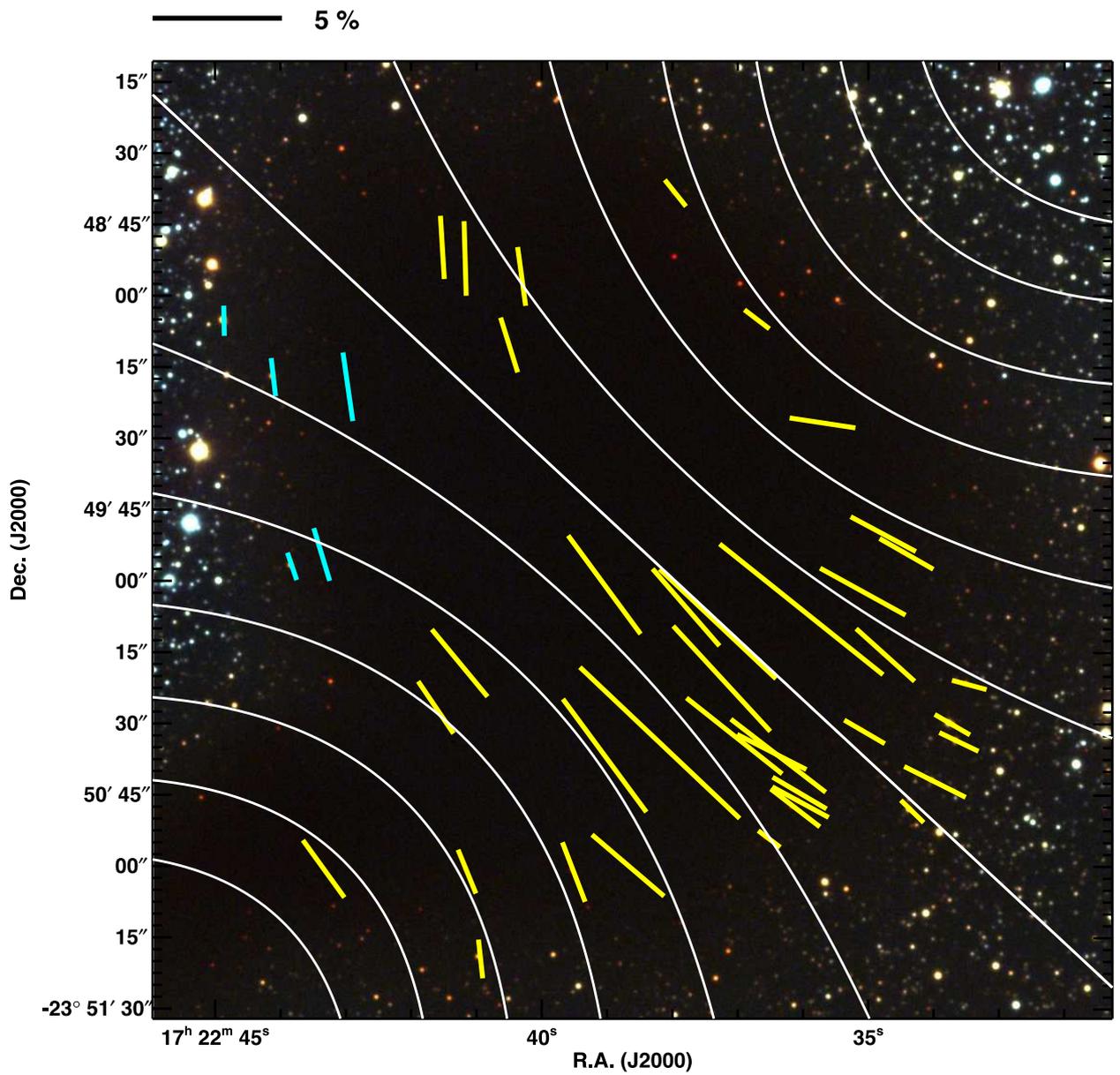}
\end{center}
 \caption{Polarization vectors after subtraction of the off-core component. The field of view is 200$''$ or 0.13 pc at a distance of 125 pc, which is equal to the diameter of the core. The background image for this figure is the optical image from Alves et al. (2001b). The white lines indicate the direction of the magnetic field inferred from parabolic fitting. The scale of the 5\% polarization degree is shown above the image.}
   \label{fig1}
\end{figure}

\clearpage 

\begin{figure}[t]  
\begin{center}
 \includegraphics[width=6.5 in]{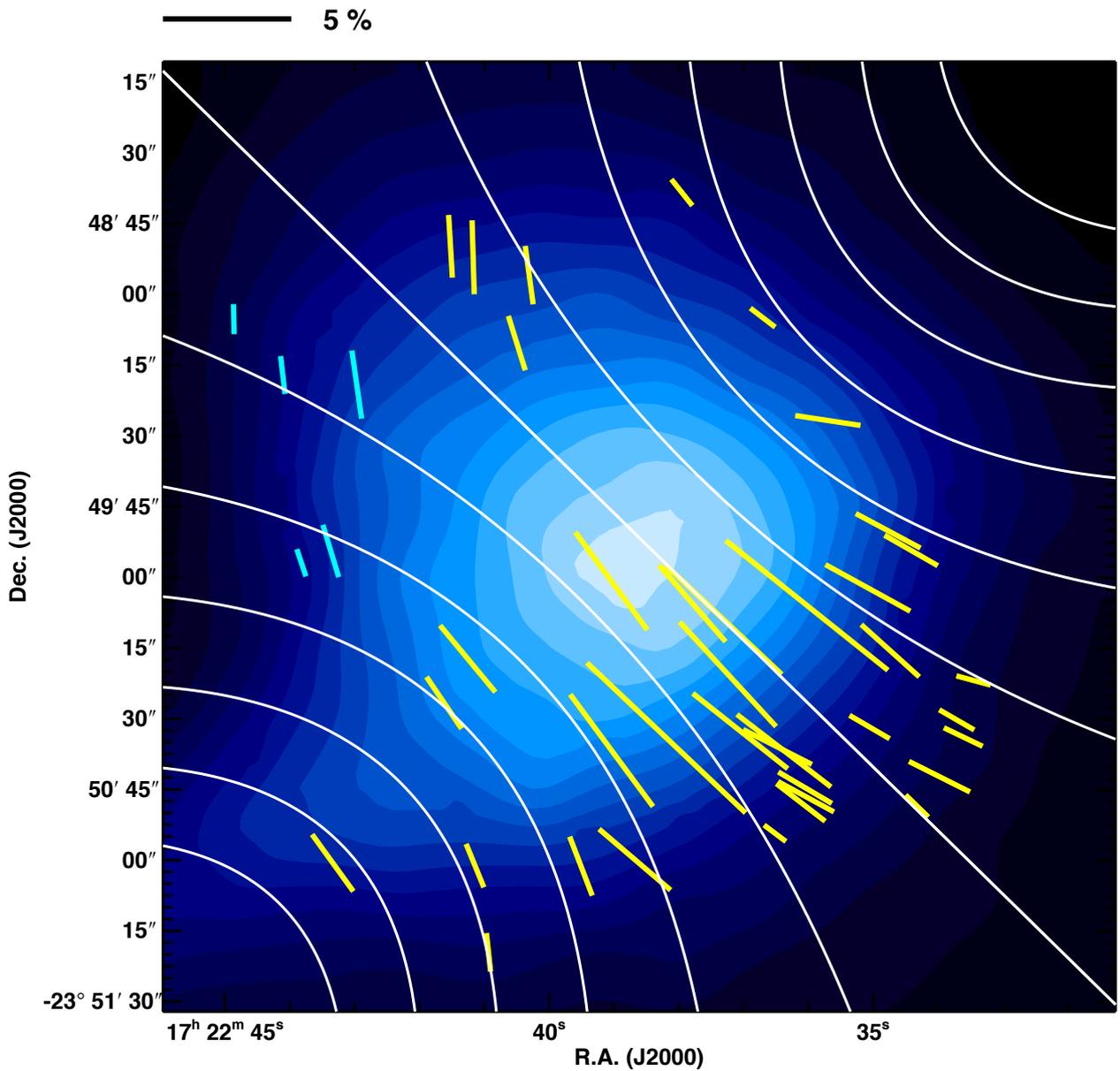}
\end{center}
 \caption{
Same as Figure 6 but the background image is a contour map of column density ($N_{\rm H_2}$) obtained with {\it Herschel} satellite (Roy et al. 2014). The contour is drawn in a step of $1.0 \times 10^{21}$ cm$^{-2}$. The minimum and maximum values are $1.6 \times 10^{21}$ and $1.0 \times 10^{22}$ cm$^{-2}$, respectively. The resolution of the map is the same as the SPIRE 500 $\mu$m data ($36.\hspace{-3pt}''3$). 
}
   \label{fig1}
\end{figure}

\clearpage 

\begin{figure}[t]  
\begin{center}
 \includegraphics[width=6.5 in]{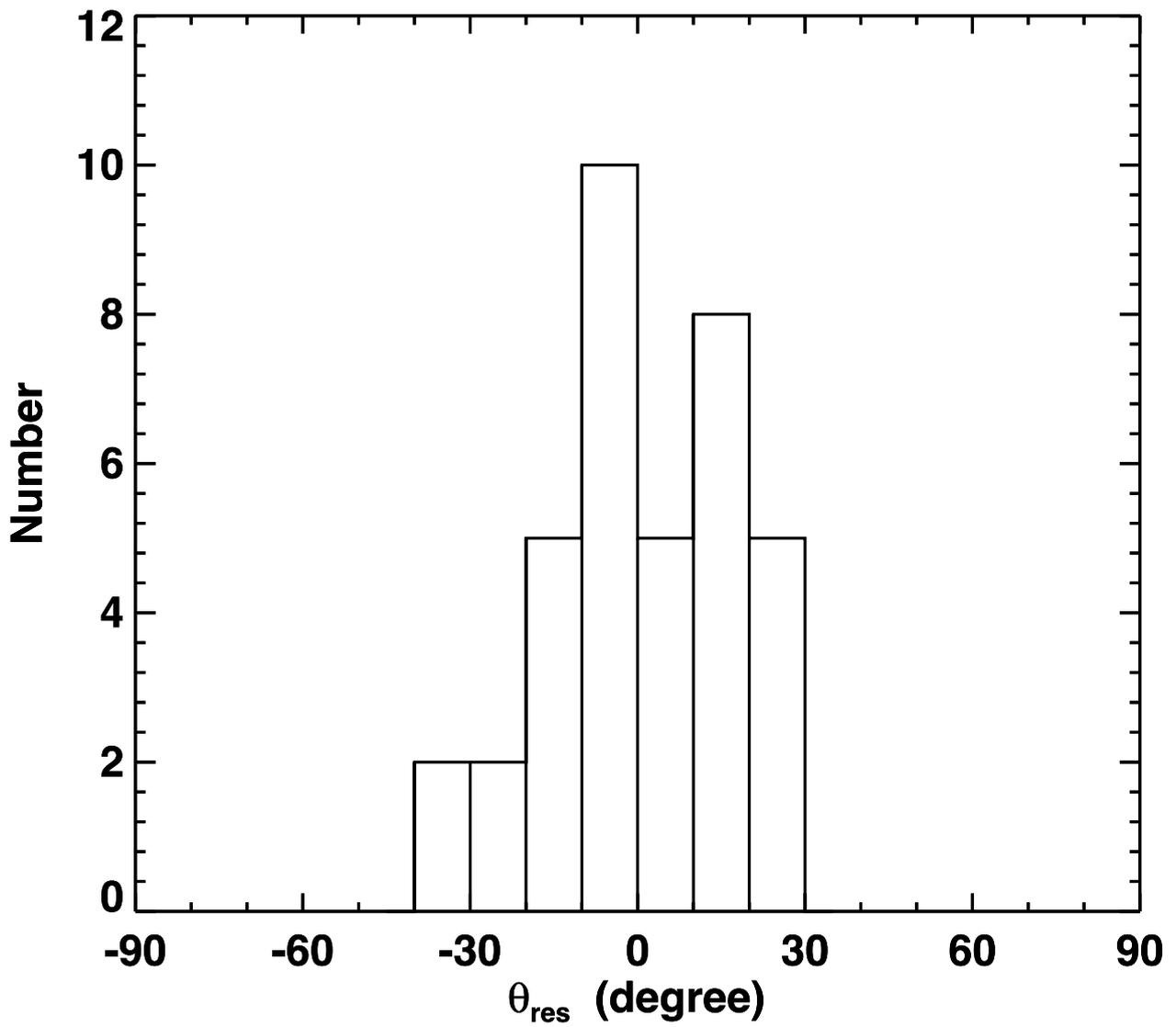}
\end{center}
 \caption{Histogram of the residual of the observed polarization angle after subtraction of the angle obtained by parabolic fitting ($\theta_{\rm res}$).}
   \label{fig1}
\end{figure}

\clearpage 

\begin{figure}[t]  
\begin{center}
 \includegraphics[width=6.0 in]{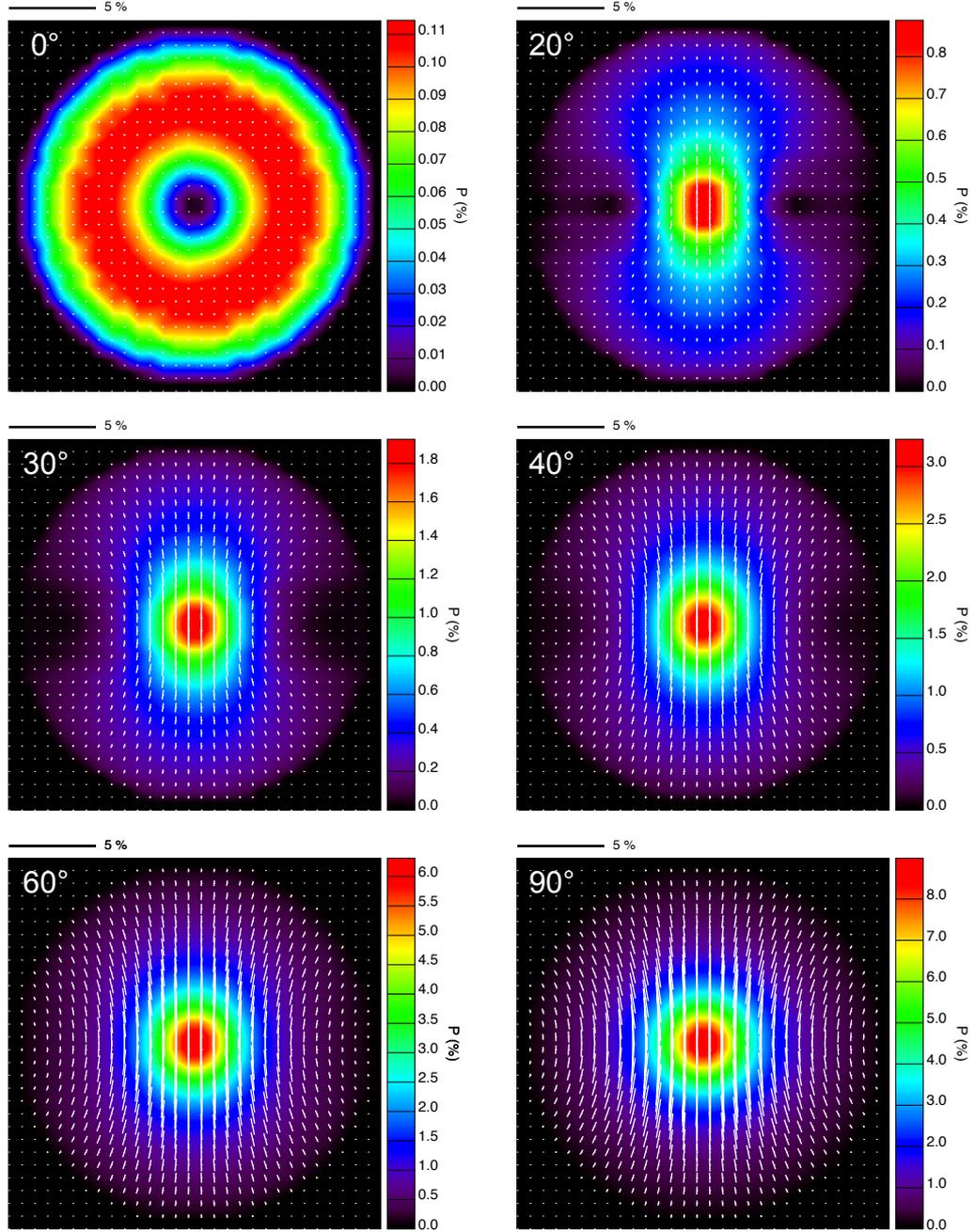}
\end{center}
 \caption{Polarization vector maps of the 3D parabolic model (white vectors). The background color and color bar show the polarization degree. The applied line-of-sight inclination angle is labeled in the upper-left corner of each panel. The 3D magnetic curvature of the model is set to $C=2.0\times10^{-4}$ ${\rm arcsec}^{-2}$ for all the panels. The scale of the 5\% polarization degree is shown above each panel.}
   \label{fig1}
\end{figure}

\clearpage 

\begin{figure}[t]  
\begin{center}
 \includegraphics[width=6.5 in]{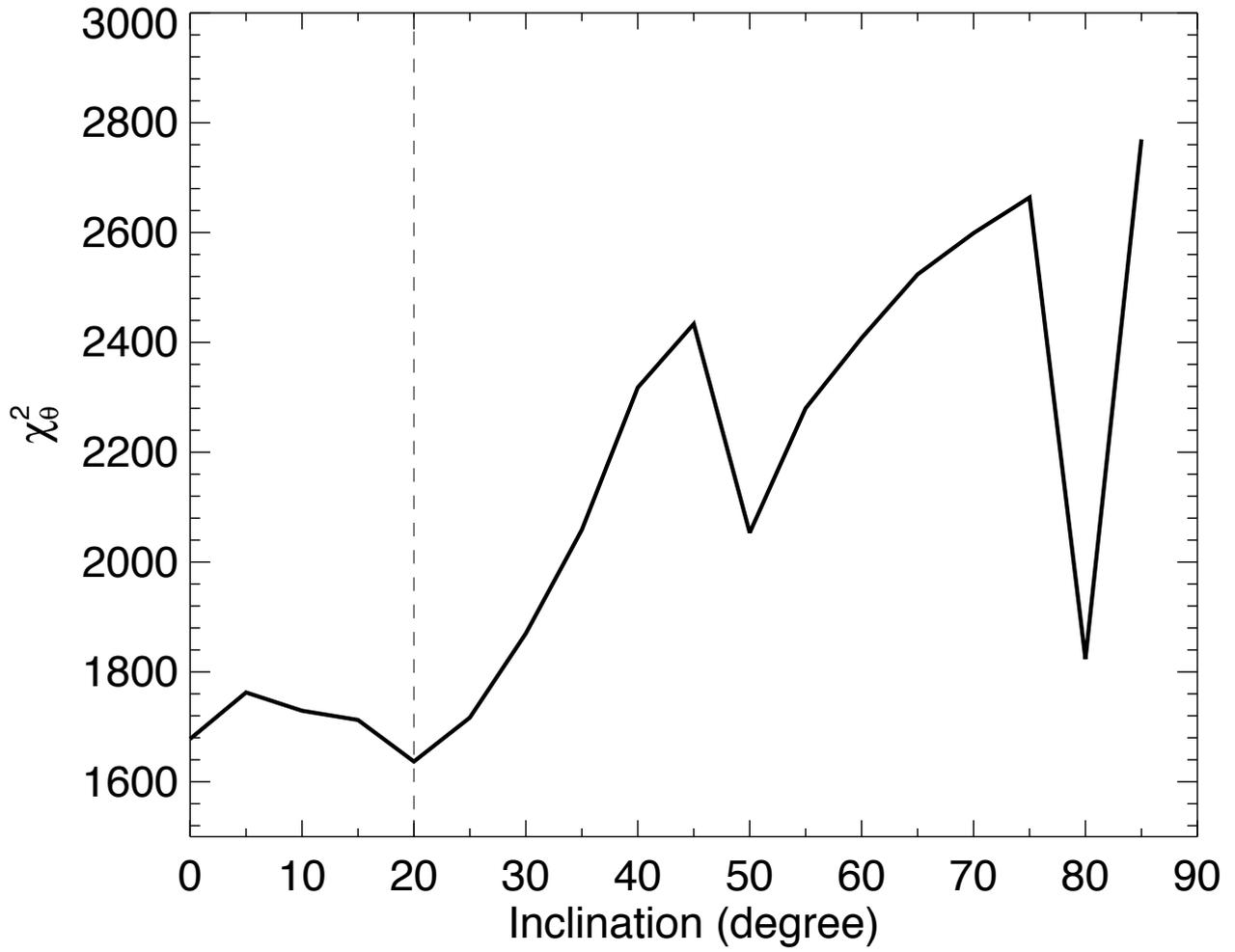}
\end{center}
 \caption{$\chi^{2}$ distribution for the polarization angle ($\chi^{2}_{\theta}$). The best magnetic curvature parameter ($C$) is determined at each $\gamma_{\rm mag}$. $\gamma_{\rm mag}=0^{\circ}$ and $90^{\circ}$ correspond to the edge-on and pole-on geometry in the magnetic axis.}
   \label{fig1}
\end{figure}

\clearpage 

\begin{figure}[t]  
\begin{center}
 \includegraphics[width=6.5 in]{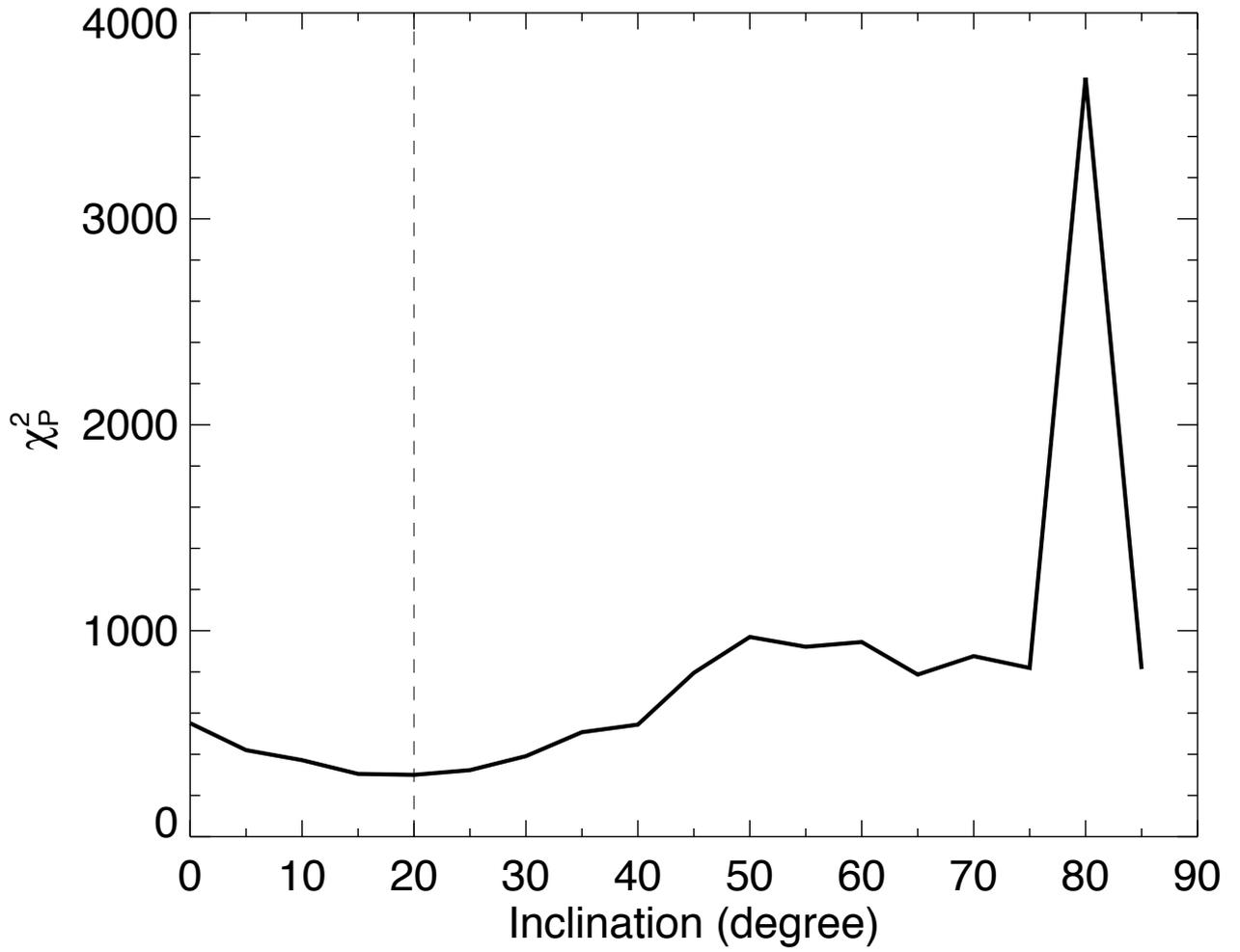}
\end{center}
 \caption{$\chi^{2}$ distribution for the polarization degree ($\chi^{2}_{P}$). $\gamma_{\rm mag}=0^{\circ}$ and $90^{\circ}$ correspond to the edge-on and pole-on geometry in the magnetic axis. Calculations of $\chi^2$ in polarization degree were performed after determining the best magnetic curvature parameter ($C$) which minimizes $\chi_2$ in the polarization angle. This calculation was carried out at each $\gamma_{\rm mag}$.}
   \label{fig1}
\end{figure}

\clearpage 

\begin{figure}[t]  
\begin{center}
 \includegraphics[width=6.5 in]{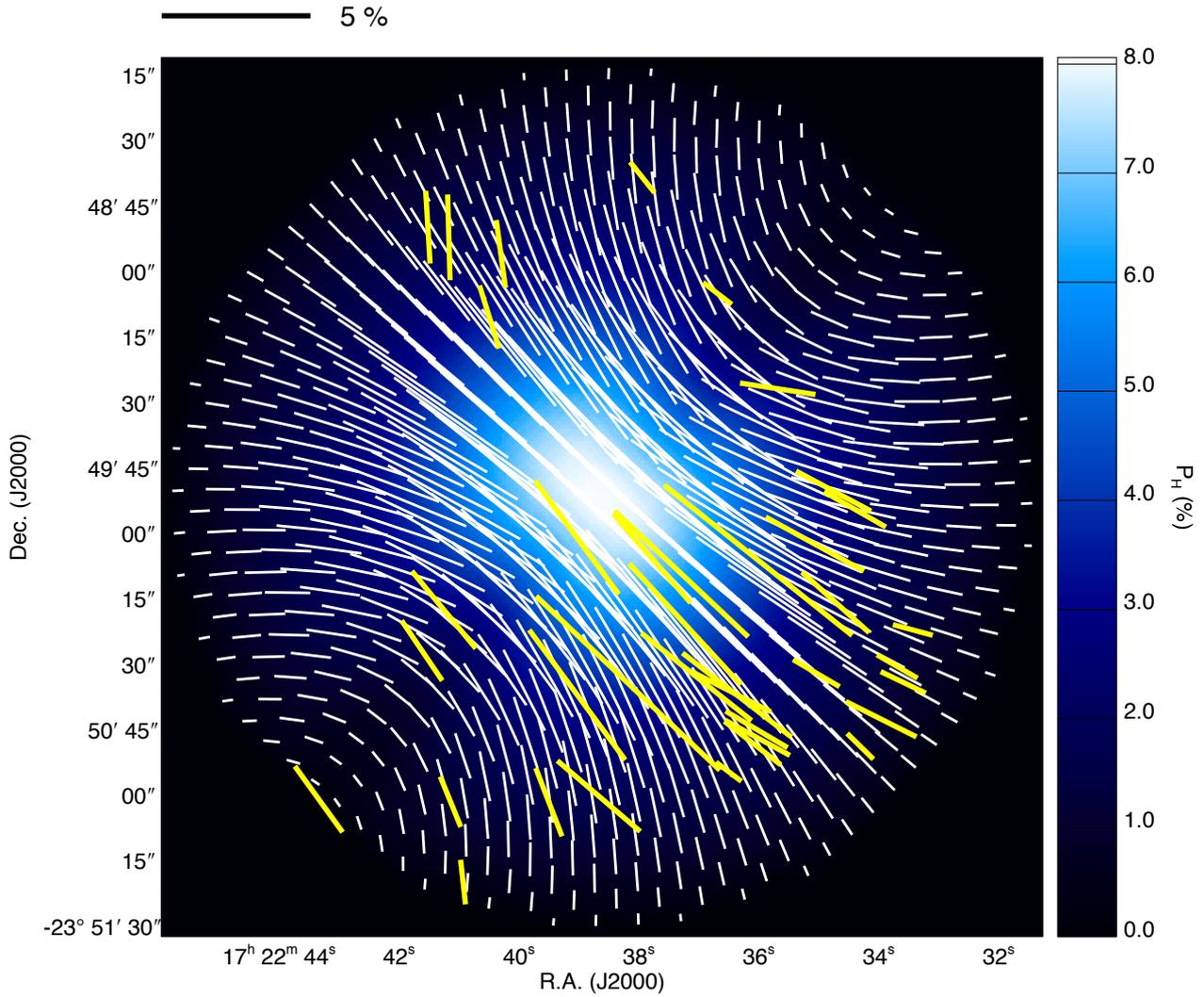}
\end{center}
 \caption{Best-fit 3D parabolic model (white vectors) with observed polarization vectors (yellow vectors). The background color image shows the polarization degree distribution of the best-fit model. The scale of the 5\% polarization degree is shown above each panel.}
   \label{fig1}
\end{figure}

\clearpage 

\begin{figure}[t]  
\begin{center}
 \includegraphics[width=6.5 in]{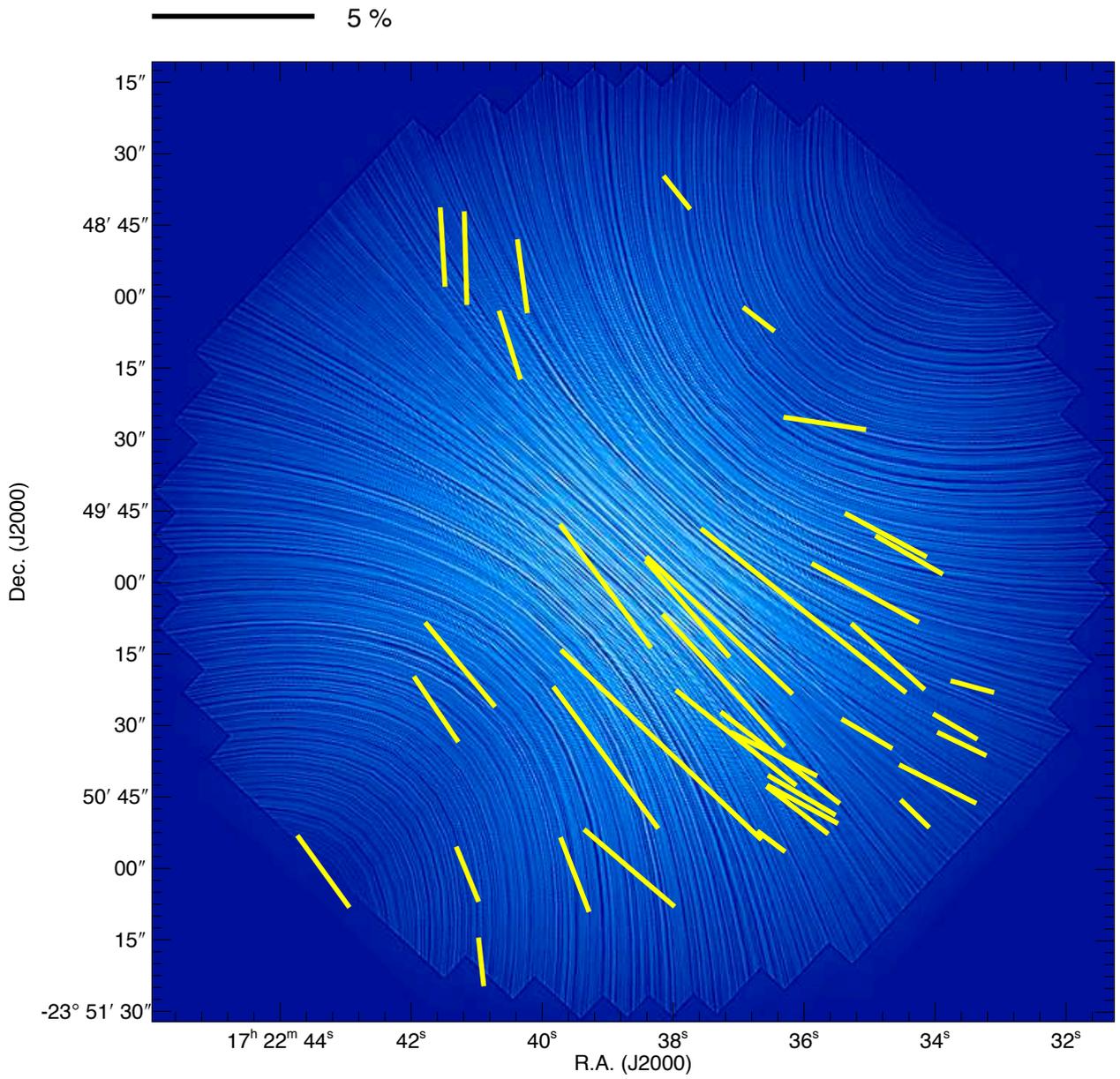}
\end{center}
 \caption{Same as Figure 12, but the background image was made using the line integral convolution technique (LIC: Cabral \& Leedom 1993). The direction of the LIC \lq \lq texture'' is parallel to the direction of magnetic fields, and the background image is based on the polarization degree of model core.}
   \label{fig1}
\end{figure}

\clearpage 

\begin{figure}[t]  
\begin{center}
 \includegraphics[width=6.5 in]{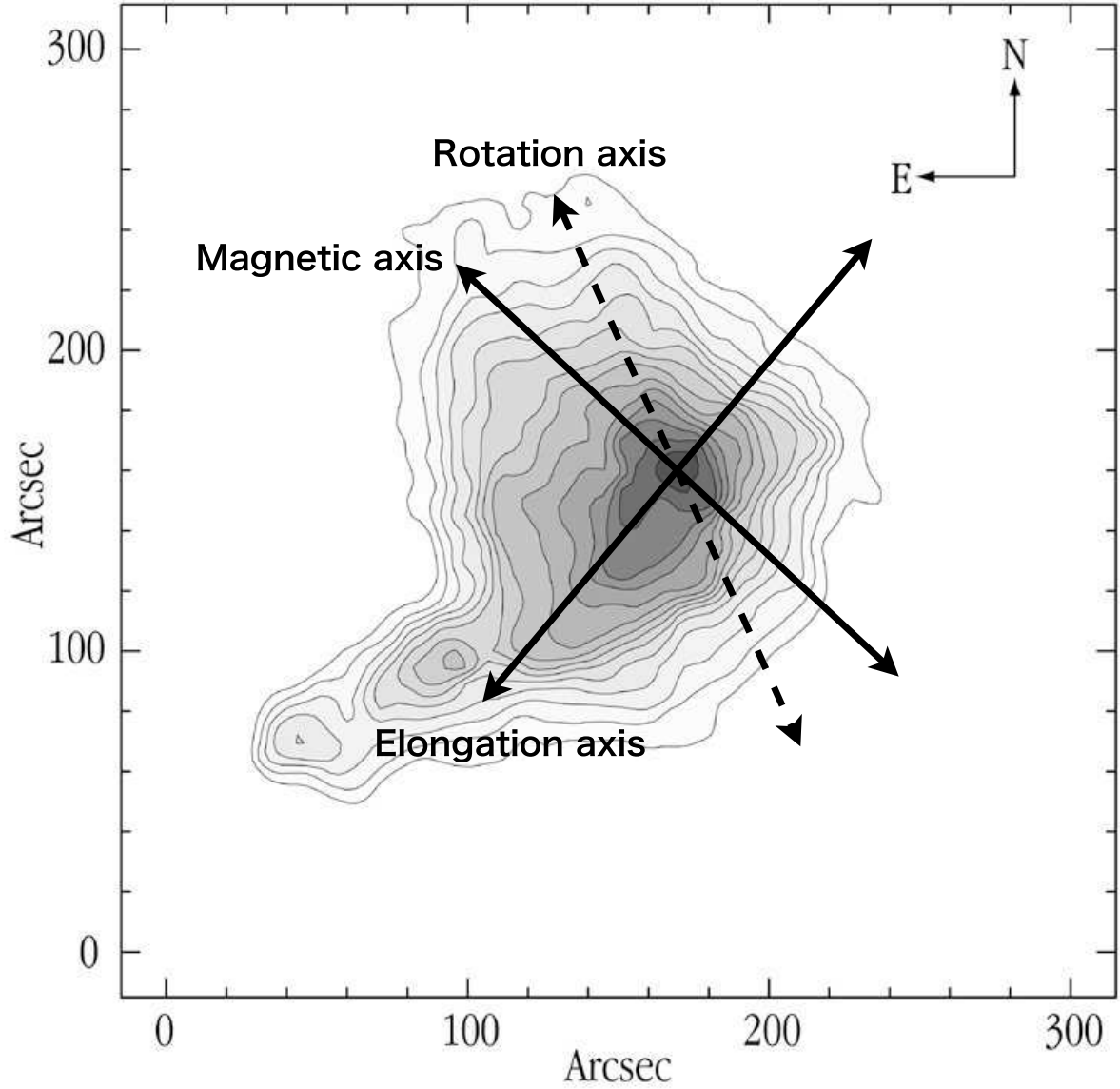}
\end{center}
 \caption{Relationship between the axis of elongation of the core ($\theta_{\rm elon} \sim 140^{\circ}$), rotation axis ($\theta_{\rm rot}=156.5^{\circ}$, Lada et al. 2003), and magnetic field axis ($\theta_{\rm mag}=47^{\circ}$) superimposed on the dust extinction ($A_V$) map taken from Alves et al. (2001a). The contours start at $A_V = 4$ mag and increase in a steps of 2 mag. The peak extinction toward the center of the core is $A_V \sim 33$ mag.}
   \label{fig1}
\end{figure}

\clearpage 

\begin{figure}[t]  
\begin{center}2
 \includegraphics[width=2.2 in]{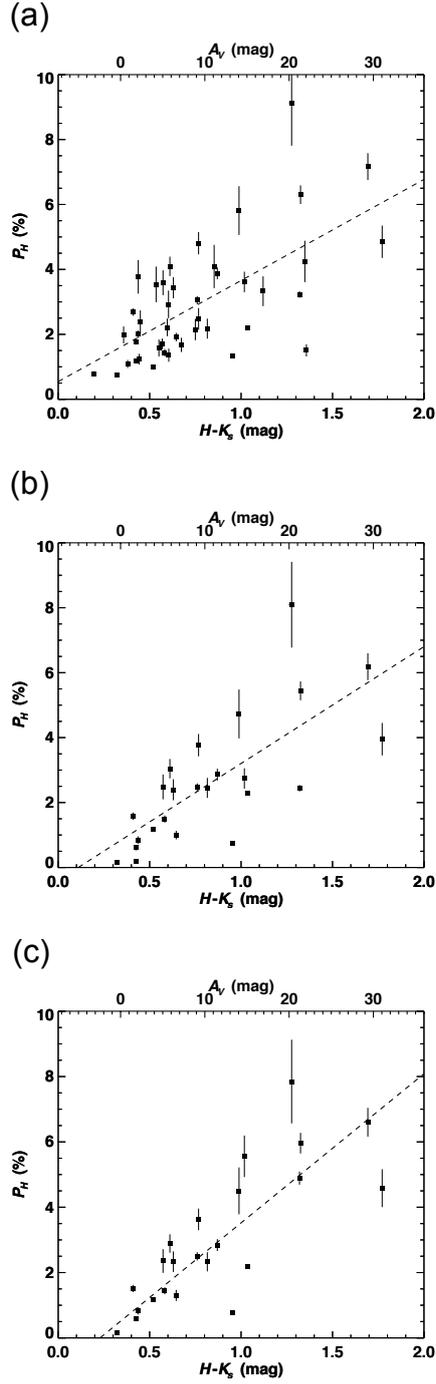}
\end{center}
 \caption{Relationship between polarization degree and $H-K_{\rm s}$ color toward background stars. The stars with $R \leq 100''$ and $P/\delta P \geq 6$ are plotted. In all the panels, the dashed lines denote the linear fit to the data. (a) $P$--$A$ relationship without any correction (original data). (b) $P$--$A$ relationship after correcting for ambient polarization components. (c) $P$--$A$ relationship after correcting for ambient polarization components, depolarization effect, and the magnetic inclination angle.}
   \label{fig1}
\end{figure}

\clearpage 

\begin{figure}[t]  
\begin{center}
 \includegraphics[width=6.5 in]{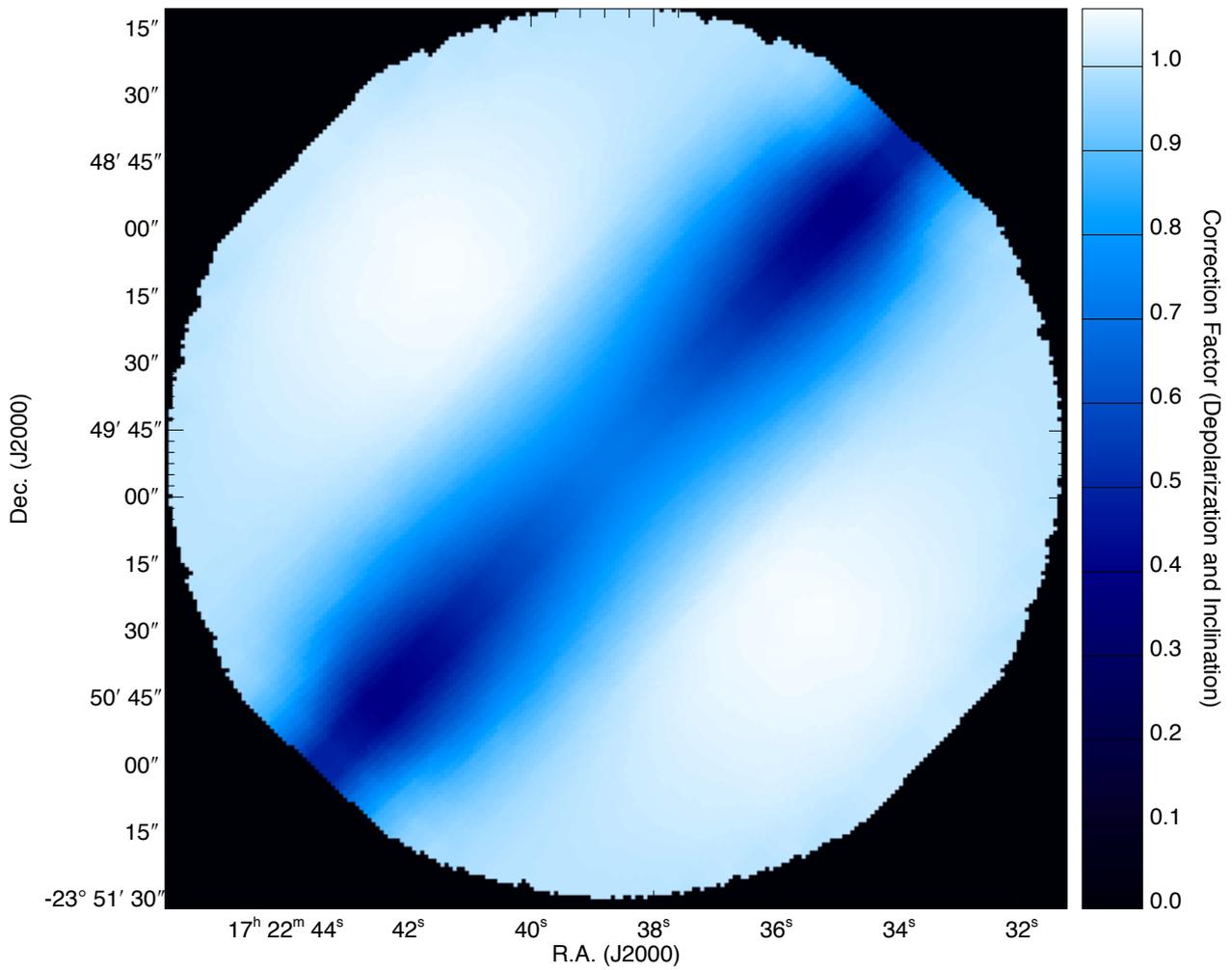}
\end{center}
 \caption{Distribution of depolarization and inclination correction factor. The field of view is the same as the diameter of the core ($200''$).}
   \label{fig1}
\end{figure}

\clearpage 

\begin{figure}[t]  
\begin{center}
 \includegraphics[width=6.5 in]{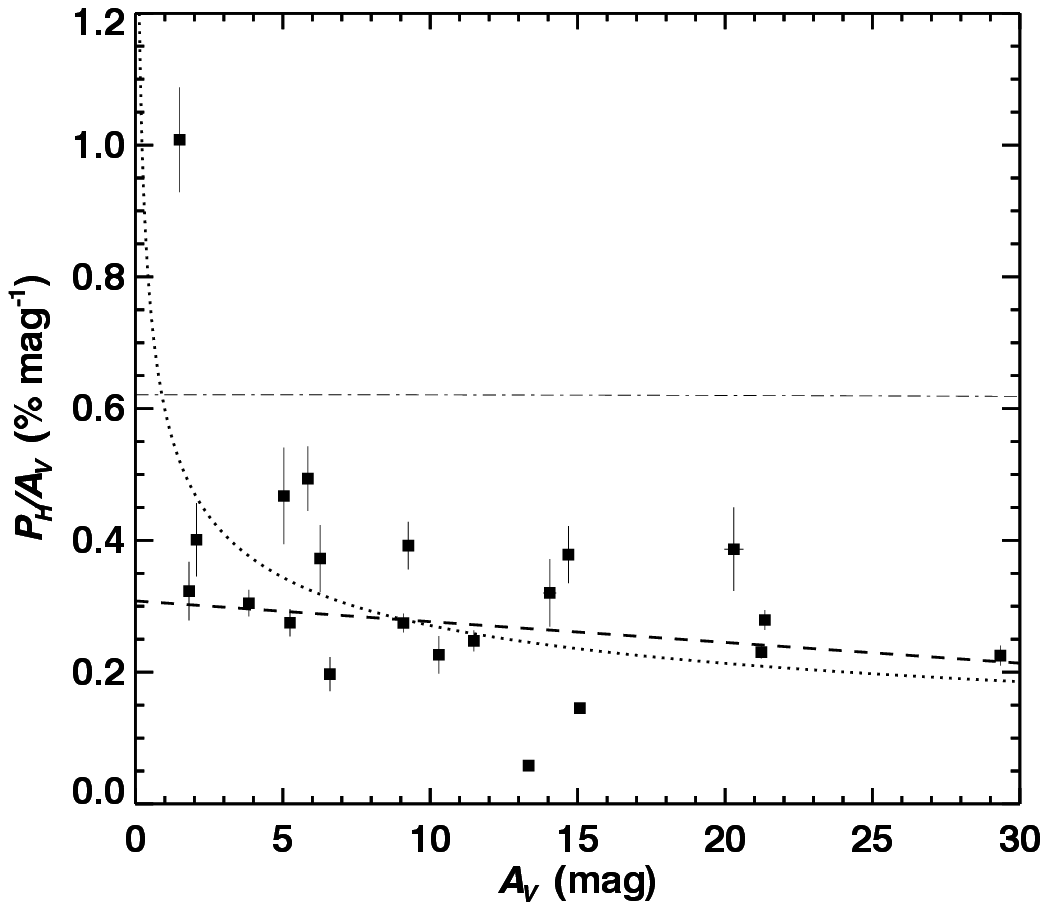}
\end{center}
 \caption{Relationship between polarization efficiency $P_H / A_V$ and $A_V$ toward background stars of B68. The stars with $R \leq 100''$ and $P/\delta P \geq 6$ are plotted. The dashed line denotes the linear fit to the data with $A_V > 7$ mag. The dotted line shows the power-law fit for whole data points. The dotted-dashed line shows the observational upper limit reported by Jones (1989).}
   \label{fig1}
\end{figure}


\begin{thebibliography}{}
\bibitem{} Alves, J. F., Lada, C. J., \& Lada, E. A., 2001, Messenger 103, 1 (Paper a) 
\bibitem{} Alves, J. F., Lada, C. J., \& Lada, E. A., 2001, Nature, 409, 159 (Paper b) 
\bibitem{} Ballesteros-Paredes, J., Klessen, R. S., \& V\'{a}zquez-Semadeni, E., 2003, ApJ, 592, 188 
\bibitem{} Bergin, E. A., Alves, J., Huard, T., \& Lada, C. J., 2002, ApJ, 570, 101 
\bibitem{} Bianchi, S., Gon\c{c}alves, J., Albrecht, M., Caselli, P., Chini, R., Galli, D., \& Walmsley, M., 2003, A\&A, 399, 43
\bibitem{} Bohlin, R. C., Savage, B. D., \& Drake J. F., ApJ, 224, 132 
\bibitem{} Bonnor, W. B., 1956, MNRAS, 116 351 
\bibitem{} Bourke, T. L., Hyland, A. R., Robinson, G., James, S. D., \& Wright, C. M., 1995, MNRAS, 276, 1067 
\bibitem{} Broderick, A. E., Keto, E., Lada, C. J., \& Narayan, R., 2007, ApJ, 671, 1832 
\bibitem{} Cabral, B. \& Leedom, L. C. 1993, in Proceedings of the 20th Annual Conference on Computer Graphics and Interactive Techniques, SIGGRAPH '93 (New York, NY, USA: ACM), 263–270
\bibitem{} Chandrasekhar, S. \& Fermi, E., 1953, ApJ, 118, 113 
\bibitem{} Davis, L., 1951, Phys. Rev., 81, 890
\bibitem{} Di Francesco, J., Hogerheijde, M. R., Welch, W. J., \& Bergin, E. A., 2002, AJ, 124, 2749 
\bibitem{} de Geus, E. J., de Zeeuw, P. T., \& Lub, J., 1989, A\&A, 216, 44 
\bibitem{} Dolginov, A. Z., \& Mitrofanov, I. G., 1976, Ap\&SS, 43, 291
\bibitem{} Draine, B. T., \& Weingartner, J. C., 1996, ApJ, 470, 551 
\bibitem{} Draine, B. T., \& Weingartner, J. C., 1997, ApJ, 480, 633 
\bibitem{} Ebert, R., 1955, ZA, 37, 217 
\bibitem{} Ewertowski, B., \& Basu, S. 2013, ApJ, 767, 33
\bibitem{} Fiedler, R. A. \& Mouschovias, T. Ch., 1993, ApJ, 415, 680 
\bibitem{} Forbrich, J., Lada, C. J., Muench, A. A., Alves, J., \& Lombardi, M., 2009, ApJ, 704, 292 
\bibitem{} Galli, D. \& Shu, F., 1993, ApJ, 417, 220 (paper a)
\bibitem{} Galli, D. \& Shu, F., 1993, ApJ, 417, 243 (paper b)
\bibitem{} Heitsch, F., Zweibel, E. G., Mac Low, M.-M., et al, 2001, ApJ, 561, 800
\bibitem{} Heitsch, F. 2005, in Astronomical Polarimetry: Current Status and Future Directions, eds. A. Adamson, C. Aspin, C. Davis, \& T. Fujiyoshi, ASP Conf. Ser., 343, 166
\bibitem{} Hildebrand, R. H., Kirby, L., Dotson, J. L., Houde, M., \& Vaillancourt, J. E., 2009, ApJ, 696, 567
\bibitem{} Hotzel, S., Harju, J., Juvela, M., Mattila, K., \& Haikala, L. K., 2002, A\&A, 391, 275 (Paper a) 
\bibitem{} Hotzel, S., Harju, J., \& Juvela, M., 2002, A\&A, 395, 5 (Paper b) 
\bibitem{} Houde, M., Vaillancourt, J. E., Hildebrand, R. H., Chitsazzadeh, S., \& Kirby, L., 2009, ApJ, 706, 1504
\bibitem{} Jones, T. J. 1989, ApJ, 346, 728 
\bibitem{} Kandori, R., Nakajima, Y., Tamura, M., et al., 2005, AJ, 130, 2166
\bibitem{} Kandori, R., Kusakabe, N., Tamura, M., et al., 2006, Proc. SPIE, 6269, 159
\bibitem{} Kandori, R., Tamura, M., Kusakabe, N., et al., 2007, PASJ, 59, 487
\bibitem{} Kandori, R., Tamura, M., Kusakabe, N., et al., 2017a, ApJ, 845, 32 (Paper I) 
\bibitem{} Kandori, R., Tamura, M., Tomisaka, K., et al., 2017b, ApJ, 848, 110 (Paper II) 
\bibitem{} Kandori, R., Tamura, M., Nagata, T., et al., 2018, ApJ, 857, 100 (Paper III) 
\bibitem{} Kandori, R., Tomisaka, K., Saito, M., et al., 2019, Submitted to ApJ (Paper VI)
\bibitem{} Kataoka, A., Machida, M., \& Tomisaka, K., 2012, ApJ, 761, 40 
\bibitem{} Keto, E., Broderick, A. E., Lada, C. J., \& Narayan, R., 2006, ApJ, 652, 1366 
\bibitem{} Lada, C. J., Bergin, E. A., Alves, J. F., \& Huard, T. L. 2003, ApJ, 586, 286 
\bibitem{} Lai, S., Velusamy, T., Langer, W. D., \& Kuiper, T. B. H., 2003, AJ, 126, 311 
\bibitem{} Lazarian, A. \& Hoang, T., 2007, MNRAS, 378, 910 
\bibitem{} Maret, S., Bergin, E. A., \& Lada, C. J., 2007, ApJL, 670, 25 
\bibitem{} Matsumoto, T., Nakazato, T., \& Tomisaka, K., 2006, ApJL, 637, 105
\bibitem{} McKee, C. F., 1989, ApJ, 345, 782 
\bibitem{} Mestel, L. \& Spitzer, L., 1956, MNRAS, 116, 503
\bibitem{} Mestel, L. 1966, MNRAS, 133, 265
\bibitem{} Mouschovias, T. Ch. \& Spitzer, L., 1976, ApJ, 210, 326  
\bibitem{} Myers, P. C., Basu, S., \& Auddy, S., 2018, ApJ, 868, 51
\bibitem{} Nagayama, T., et al., 2003, Proc. SPIE, 4841, 459 
\bibitem{} Nakano, T. \& Nakamura, T., 1978, PASJ, 30, 671 
\bibitem{} Nielbock, M., et al., 2012, A\&A, 547, 11
\bibitem{} Nishiyama, S., Nagata, T., Tamura, M., Kandori, R., Hatano, H., Sato, S., \& Sugitani, K., 2008, ApJ, 680, 1174
\bibitem{} Ostriker, E. C., Stone, J. M. \& Gammie, C. F., 2001, ApJ, 546, 980 
\bibitem{} Padoan, P., Goodman, A., Draine, B. T., Juvela, M., Nordlund, \r{A}., \& R\"{o}gnvaldsson, \"{O}. E., 2001, ApJ, 559, 1005
\bibitem{} Redman, M. P., Keto, E., \& Rawlings, J. M. C., 2006, MNRAS, 370, 1 
\bibitem{} Roy, A., Andr\'{e}, P., Palmeirim, P., et al., 2014, A\&A, 562, 138 
\bibitem{} Tomisaka, K., Ikeuchi, S., \& Nakamura, T., 1988, ApJ, 335, 239 
\bibitem{} Wardle, J. F. C., \& Kronberg, P. P. 1974, ApJ, 194, 249
\end{thebibliography}
\end{document}